\documentclass[entropy,article,accept,moreauthors,pdftex,10pt,a4paper]{mdpi}
\usepackage{soul}
\usepackage{upgreek}
\usepackage{enumitem}
\usepackage{subfigure}
\usepackage{booktabs}
\usepackage{multirow}
\usepackage{longtable}
\usepackage{microtype}
\newcommand\myurl[1]{\changeurlcolor{black}\url{#1}\changeurlcolor{blue}}

\firstpage{1}
\articlenumber{88}
\doinum{10.3390/e19030088
}
\pubvolume{19}
\pubyear{2017}
\copyrightyear{2017}
\externaleditor{Academic Editor: Geert Verdoolaege}
\history{Received: 4 February 2017; Accepted: 22 February 2017; Published: 24 February 2017}
\graphicspath{{./figures/}}
\usepackage{soul}

\pdfoutput=1

\usepackage{amssymb}
\usepackage{multirow}

\newcommand{\dd}{ {\textrm d}}
\newcommand{\ee}{ {\textrm e}}
\Title{Systematic Analysis of the Non-Extensive Statistical~Approach in High Energy Particle Collisions---Experiment vs. Theory $^{\dagger}$}

\Author{G\'abor B\'ir\'o $^{1,2,}$*, Gergely G\'abor Barnaf\"oldi $^{1}$, Tam\'as S\'andor Bir\'o $^{1}$, K\'aroly \"Urm\"ossy $^{3}$ and~\mbox{\'Ad\'am Tak\'acs $^{2}$}}
\AuthorNames{G\'abor B\'ir\'o, Gergely G\'abor Barnaf\"oldi, Tam\'as S\'andor Bir\'o, K\'aroly \"Urm\"ossy and \'Ad\'am Tak\'acs}

\address{%
$^{1}$ \quad Wigner Research Center for Physics of the HAS, 29--33 Konkoly--Thege Mikl\'os Str., H-1121 Budapest, Hungary; barnafoldi.gergely@wigner.mta.hu (G.G.B.); biro.tamas@wigner.mta.hu (T.S.B.)\\
$^{2}$ \quad Institute of Physics, E\"otv\"os Lor\'and University, 1/A P\'azm\'any P\'eter S\'et\'any, H-1117  Budapest, Hungary; takacs.adam@wigner.mta.hu\\ 
$^{3}$ \quad Institute of Physics, Jan Kochanowski University, 5 \.Zeromskiego St., 25-369 Kielce, Poland; urmossy.karoly@wigner.mta.hu} 

\corres{Correspondence: biro.gabor@wigner.mta.hu; Tel.: +36-30-308-6742}

\abstract{
The~analysis of high-energy particle collisions is an~excellent testbed for the non-extensive statistical approach. In these reactions we are far from the thermodynamical limit.
In small colliding systems, such as electron-positron or nuclear collisions, the~number of particles is several orders of magnitude smaller than the Avogadro number; therefore, finite-size and fluctuation effects strongly influence the final-state one-particle energy distributions. Due to the simple characterization, the~description of the identified hadron spectra with the Boltzmann--Gibbs thermodynamical approach is insufficient. 
These spectra can be described very well with Tsallis--Pareto distributions instead, derived from non-extensive thermodynamics. 
Using the $q$-entropy formula, we~interpret the microscopic physics in terms of the Tsallis $q$ and $T$ parameters.
In this paper we give a~view on these parameters, analyzing identified hadron spectra from recent years in a~wide center-of-mass energy range. We demonstrate that the fitted Tsallis-parameters show dependency on the center-of-mass energy and particle species (mass). Our findings are described well by a~QCD (Quantum Chromodynamics) inspired parton evolution ansatz.
Based on this comprehensive study, apart from the evolution, both mesonic and baryonic components found to be non-extensive ($q>1$), besides the mass ordered hierarchy observed in the parameter $T$.
We also study and compare in details the theory-obtained parameters for the case of PYTHIA8 Monte Carlo Generator,
perturbative QCD and quark coalescence models.}


\conference{MaxEnt 2016}

\conferencetitle{MaxEnt 2016}

\begin{document}


\section{Introduction}

In Nature we often meet phenomena with a~large number of variables where the few-body approach breaks down. In these cases the standard procedure is to apply tools from statistical physics and inspect thermodynamical quantities of the system, instead of treating all degrees of freedom one-by-one. A certain generalization of the standard Boltzmann--Gibbs entropy is promoted by Constantino Tsallis, introducing the $q$-entropy formula, central to non-extensive statistical theory~\cite{book:ts, book:ts2}. Despite its unconventional form, in~the last two decades the Tsallis-entropy was found to be a~very general and descriptive notion. Numerous physical observations were successfully explained using non-extensive statistical physics~\cite{book:nonextcollection, TsallisBibHtml, artic:tsappl1, artic:tsappl2, artic:tsappl3}. From our perspective the field of high-energy physics is especially important, since that community uses efficiently these tools to describe the results of high-energy particle and nuclear collisions. It is an~experimental finding that the distributions derived from Tsallis-entropy fit the spectrum of high-energy particles, produced by many systems starting from electron-positron collisions up to the cosmic rays. In this paper we focus on identified hadron spectra, measured in proton-proton collisions. We put emphasis on the investigation of the center-of-mass energy ($\sqrt{s}$) dependence of the Tsallis parameters $q$ and $T$, assuming a~Quantum Chromodynamics (QCD) inspired logarithmic scaling of these parameters. We use units in which $\hbar=c=k_B=1$.

The~outline of the paper is the following: in the next section we enlist our motivation from high energy physics and the goals of our analysis. In Section~\ref{sec:second} we briefly introduce the mathematical apparatus we used during our investigation. In Section~\ref{sec:third} we show experimental results, while in Section~\ref{sec:fourth} we compare them to state-of-the-art theoretical models. Finally, in~Section \ref{sec:fifth} we summarize our work and give a~discussion, including our future plans.

\section{Connection with High Energy Physics}
\label{sec:first}

One of the main goals in high-energy heavy-ion physics is to understand the properties of the so-called Quark Gluon Plasma (QGP), a~particular form of the strongly interacting matter which existed shortly after the Big Bang. With today's high-energy particle accelerators we are able to reach the energy range where this superdense matter of the early Universe can be formed for a~short, $\mathcal{O}$(fm/$c$)$\sim10^{-23}$ s time. The~properties of the QGP can be studied in ultra-relativistic heavy-ion collisions indirectly. Due to the nature of the strong interaction there is no way for direct observation, only signatures stemming from the final state allow us to draw conclusions.
On the other hand, the~reactions occur during a~very short time and our information about their nature is very limited. This is a~strong restraint in our possibilities, especially we cannot treat properly the description at the microscopical level. Nevertheless it is essential to understand the processes in proton-proton collisions, the~baseline for heavy-ion measurements.

To date we still do not have a~well established, detailed, and~throughout probed theory of the {\em hadronization}, the~process where the color degrees of freedom confine into hadrons. This is related to the {\em Yang--Mills Mass Gap}, one of the so-called ``Millennium problems'' of the Clay Mathematical Institute~\cite{artic:millenium}. Recent hadronization models are phenomenological, and~it is quite typical that their parameters lack of any clear physical meaning.

Recently, complex detector systems, like ALICE at the Large Hadron Collider (CERN LHC) or STAR and PHENIX at the Relativistic Heavy Ion Collider (BNL RHIC) are able to measure with high accuracy the final state particles.
The~hadron spectra, measured in high-energy collisions, are one of the most fundamental characteristics of these events and involve both microscopic and collective effects in high-energy collisions. Identifying their most crucial problems is a~key task for understanding hadronization. A remarkable phenomenon is that these properties occur not only in heavy-ion collisions, but even for small colliding systems like proton-proton or electron-positron collisions~\cite{artic:flowing0, artic:flowing1, artic:flowing2, artic:flowing3, artic:gbiro}.

The~aim of our study is to find the common source of these similarities and to recognize the driving mechanisms behind the observations. For this aim, we~built a~consistent non-extensive approach, in~which fit parameters carry important physical information about the observed system of high-energy and strongly interacting particles including reactions and collective effects among them.

\section{Non-Extensive Statistics in High-Energy Physics}
\label{sec:second}

High-energy physics, and~in particular high-energy heavy ion physics is an~interdisciplinary topic. It uses the theory of relativistic quantum fields, statistical physics, thermo- and hydrodynamics, and~even the theory of curved space-times. Earlier studies show that non-extensive statistical physics provides a~useful tool to describe particle-particle collisions, where ``{\it particle}'' now stands either for electron/positron, or for proton or a~heavy nucleus. The~non-extensivity in high-energy physics manifests itself both in the non-exponential energy- and non-Poissonian multiplicity distributions.

The~hadron spectra can be characterized with Tsallis--Pareto-like distributions, both at low-, and~high transverse momenta very well~\cite{artic:gbiro, artic:tsbeurphys2,  artic:cleymansphyslett, artic:cleymansjphys, artic:wilk1, artic:wilk2, artic:wilk3, artic:wilk4, artic:tsorig, artic:wilkentr, artic:actaphys, artic:chaossolitions}.
The~origin of these distributions lies in the assumption that subsystems are not independent of each other. It makes them a~good candidate to investigate the QGP in heavy ion collisions or the baseline in smaller colliding systems. Recently a~number of systematic analyses have been made in order to find the best form for these distributions~\cite{artic:lilin1, artic:lilin2, artic:cleymansTS, artic:cleymansqspecies, artic:flowing0, artic:flowing1, artic:flowing2, artic:flowing3}, but in this present study we compare theoretical and experimental results in a~wide energy range comprehensively. 

\subsection{The~Description of the Inclusive Hadron Production}
\label{sec:hadronprod}

The~Quantum Chromodynamics is the fundamental theory of the strong interaction. Due to the energy-scale dependent behavior of the strong coupling, the~perturbative QCD (pQCD) based parton model--initiated by Bjorken and Feynman--works extremely well at high energies~\cite{artic:bjorken}. In the framework of the pQCD-based parton model, all hadrons are made up from partons (bare, nearly massless quarks and gluons), therefore the inner structure of the initial colliding and the finally produced hadrons are described by the parton distribution functions (PDF) and by the fragmentation functions (FF), respectively. These~non-perturbative distribution functions are defined in the momentum space and can be parametrized by a~polynomial {\em ansatz}. The~PDF,
\begin{equation}
f_{a/h}(x_a, Q^2) \ \ ,
\end{equation}
gives the distribution of parton $a$ inside the hadron $h$ at the energy scale, $Q^2$,  while $x_a=p_a/p_h$ is the momentum fraction carried by that parton. On the other hand, the~confinement of the parton $c$ into the final state hadron $h$ with the momentum fraction $z_c=p_h/p_c$ can be described at scale, $\tilde{Q}^2$ with the help of fragmentation functions
\begin{equation}
D_c^h(z_c,\tilde{Q}^2) .
\end{equation}

In this framework the inclusive cross-section of a~given hadron $h$ produced in proton-proton collisions can be calculated by the following convolution:
\begin{equation}
\label{eq:hadronprod}
E_h\frac{\dd^3 \sigma_h^{pp}}{\dd p_h^3}\sim \sum\limits_{a,b,c}f_{a/p}(x_a,Q^2)\otimes f_{b/p}(x_b,Q^2) \otimes \frac{\dd \sigma^{ab\rightarrow cd}}{\dd \hat{t}} \otimes \frac{D_c^h(z_c,\tilde{Q^2})}{\pi z_c^2} \ \ \ .
\end{equation}

Here the parton distribution function of a~proton is denoted by $f_{a/p}(x_a, Q^2)$ and $\frac{\dd \sigma}{\dd \hat{t}}$ is the differential cross-section of the $ab\rightarrow cd$ partonic process, the~variable $\hat{t}$ is related to the 4-momentum exchange of the particles.

The~hadronization is described within the parton model by the above phenomenological fragmentation functions, for which several forms of parametrization exist in the literature. These~parametrizations are usually fitted to lepton scattering data, therefore they describe existing experimental results in a~broad-range in the parameter space. In Section \ref{sec:fourth}, after investigating the energy dependence, we~show the latest results of a~new fragmentation function parametrization based on non-extensive phenomena.

\subsection{Hadronization Using Non-Extensive Statistics}
\label{sec:non-exthadron}

As we have already mentioned, the~transverse energy distribution of the measured hadrons---the particle yield measured in the $y \in [-0.5,+0.5] $ midrapidity region--is an~important  quantity accessible to measurement. In practice, the~low-energy regime is described by exponential-like functions, as~a~thermalized system, while the high $p_T$ regime behaves like a~power-law, $p_T^{-n}$. The~Tsallis--Pareto-like distributions handle these two regimes simultaneously.

The~technical apparatus in the high-energy physics shows a~great advancement, nowadays the statistics of these spectra is larger than ever. It is no surprise that the Tsallis--Pareto distributions are widely used by the high-energy community to describe hadron spectra. The~STAR and PHENIX collaborations at RHIC BNL (USA) and the European CERN's ALICE, ATLAS, and~CMS collaborations at the LHC are using the following form to characterize the particle yield~\cite{artic:phenixts, artic:starts, artic:62gevphenix, artic:097tevalice, artic:cmsts, artic:atlasts}:
\begin{equation}
\frac{1}{2\pi p_T}\frac{\dd^2N}{\dd y \dd p_T} = \frac{\dd N}{\dd y}\frac{(n-1)(n-2)}{2\pi n C\left(nC+m(n-2)\right)}\left(1+\frac{m_T-m}{nC}\right)^{-n} \ \ \ ,
\label{eq:tsexp}
\end{equation}
where $n$ and $C$ are fit parameters and $m_T=\sqrt{p_T^2+m^2}$ is the transverse mass, including the rest mass $m$ of the given identified hadron species. We note that this formula is based on the QCD-Hagedorn formula~\cite{artic:qcdhagedorn, artic:ua1hagedorn1, artic:ua1hagedorn2, artic:mpprod1, artic:mpprod2}. This and other variations of the distribution are exhaustively tested e.g.,~in~\cite{artic:gbiro, artic:tsbeurphys2, artic:flowing0, artic:flowing1, artic:flowing2, artic:flowing3, artic:cleymansphyslett, artic:lilin1, artic:lilin2}. Below we theoretize over the origin of such Tsallis-type formulas. Contrary to the fixed fit parameters of the Tsallis--Pareto distributions as in Equation~(\ref{eq:tsexp}), we~assume that the identified hadron spectra are characterized with a~scaling Tsallis-distribution, where an~energy scaling of the Tsallis-parameters is also present. In the following we refer to these as Tsallis-like~distributions.

In extensive systems the entropy is finite in the thermodynamical limit, $\lim\limits_{N\rightarrow\infty}\frac{S_N}{N}<\infty$. This is the case with the Boltzmann--Gibbs--Shanon entropy formula, $S=-\sum\limits_i P_i \ln{P_i}$, where $P_i$ is the probability of being in state $i$. In strongly correlated systems, it turns out that the total entropy of the system is not the sum of the entropy of the subsystems:
\begin{equation}
S_{12}\neq S_1 + S_2 \ \ .
\end{equation}

For our generalization, we~use the well established terminology of the thermodynamics, since we expect to include the classical Boltzmann--Gibbs case too. Let us consider a~monotonic, transformed entropy, $L(S)$, which satisfies additivity ,
\begin{equation}
L(S_{12})=L(S_1)+L(S_2) \ .
\end{equation}
Note, on general terms $L(S)$ is the logarithm of the formal group of phase space factors $\Omega(S)=e^S$.
Due to this assumption, applied recursively in ensembles, we~arrive at the following general class of entropies~\cite{artic:tsbentr}:
\begin{equation}
L(S)=\sum\limits_i P_i L(- \ln{P_i}) \ \ \ .
\label{eq:entrdef}
\end{equation}

Because $L(S)$ is by definition a~monotonic function, the~most likely state of a~heat reservoir and its subsystem at maximum entropy is equivalently at
\begin{equation}
\label{eq:l-maxent}
L\left( S_1(E_1^* \right) )+L \left( S_2(E_2^*) \right)= \mathrm{max} \ \ \ ,
\end{equation}
where $E_1$ is the energy of the subsystem, $E_2=E-E_1$ is the energy of the reservoir. While keeping $E~=~$const in the entropy maximum Equation~(\ref{eq:l-maxent}) we obtain
\begin{equation}
0= \left.\frac{\partial L}{\partial S_1}\right|_{S_1(E_1^*)} \times \left.\frac{\partial S_1}{\partial E_1}\right|_{E_1^*}
	-\left.\frac{\partial L}{\partial S_2}\right|_{S_2(E_2^*)} \times \left.\frac{\partial S_2}{\partial E_2}\right|_{E_2^*} \ \ \ .
\end{equation}

It makes the use of the usual definition of thermodynamical temperature expedient,
\begin{equation}
\beta_1 :=\left.\frac{\partial L}{\partial S_1}\right|_{S_1(E_1^*)} \times \left.\frac{\partial S_1}{\partial E_1}\right|_{E_1^*} =
            \left.\frac{\partial L}{\partial S_2}\right|_{S_2(E_2^*)} \times \left.\frac{\partial S_2}{\partial E_2}\right|_{E_2^*} = \beta_2 \ \ \ .
            \label{eq:beta1}
\end{equation}

Assuming now $E_1\ll E$ in high-energy collisions, we~consider: $E\rightarrow\sqrt{s}$ and $E_1\rightarrow (m_T-m)\approx p_T$ of the particle,
\vspace{12pt}
\begin{equation}
L(S_2(E-E_1))\approx L\left(S_2(E)-\left.\frac{\partial S_2}{\partial E_2}\right|_E \times E_1\right) \approx L(S_2(E)) -
	\left.\frac{\partial L}{\partial S_2}\right|_{S_2(E)} \times \left.\frac{\partial S_2}{\partial E_2}\right|_{E} \times E_1 \ \ .
\end{equation}

Inserting this into Equation~(\ref{eq:beta1}), we~arrive at the formula,
\begin{equation}
\beta_1 \approx  \left.\frac{\partial L}{\partial S_2}\right|_{S_2(E)} \times \left.\frac{\partial S_2}{\partial E_2}\right|_{E}
 - \left[
\left.\frac{\partial^2 L}{\partial S^2_2}\right|_{S_2(E)} \times \left(\left.\frac{\partial S_2}{\partial E_2}\right|_{E} \right)^2 +
	  \left.\frac{\partial L}{\partial S_2}\right|_{S_2(E)} \times \left.\frac{\partial^2 S_2}{\partial E^2_2}\right|_{E} \right] \times E_1 + \dots \ \ .
	  \label{eq:beta1_2}
\end{equation}

By looking for an~{\em universal termostat}, lending to $\beta_1$ an~absolute temperature interpretation, we~assume that the energy of the subsystem is independent from the energy of the reservoir, i.e.,~we~require the term linear in $E_1$ to vanish. After ordering we obtain:
\begin{equation}
\left.\left.\frac{\partial^2 L}{\partial S^2_2}\right|_{S_{2}(E)}\right/ \left.\frac{\partial L}{\partial S_2}\right|_{S_2(E)} =
	 - \left.\left.\frac{\partial^2 S_2}{\partial E^2_2}\right|_{E}\right/ \left(\left.\frac{\partial S_2}{\partial E_2}\right|_{E}\right)^2 \ \ \ .
\end{equation}

This equality among general functions, $L(S)$ and $S(E)$ is possible only if both are equal with a~constant,
\begin{equation}
\frac{L''(S)}{L'(S)} = - \frac{S''(E)}{S'(E)^2} =\frac{1}{C} := 1-q \ ,
\end{equation}
where $C$ is the heat capacity of the reservoir.

The~solution of this differential equation has all desired features:
\begin{equation}
L(S)=\frac{\ee^{(1-q)S}-1}{1-q} \ \ \ .
\end{equation}

Replacing it into the Equation~(\ref{eq:entrdef}),
\begin{equation}
L(S)= \sum\limits_i P_i \frac{\ee^{-(1-q) \ln P_i}-1}{1-q} = \frac{1}{1-q} \sum\limits_i
P_i \left( P_i^{q-1} -1\right) = \frac{1}{1-q} \sum\limits_i
 \left( P_i^{q} -P_i \right)
\end{equation}
is the (now additive) Tsallis entropy, while
\begin{equation}
S= \frac{1}{1-q} \ln \left( 1+ (1-q) L(S) \right) = \frac{1}{1-q} \ln \sum\limits_i P_i^q \  \ ,
\end{equation}
turns out to be the R\'enyi entropy.

This argumentation can be used also for microcanonical systems, with $S_1=- \ln{P_1}$ and $P_1$ being the distribution of the subsystem's states. Using the previously-defined generalized  entropy, $L(S)$, one~arrives at the following energy distribution, which maximizes the $q$-entropy:
\begin{equation}
P_i = \left( Z^{1-q} + (1-q)\frac{E_i}{T}\right)^{-\frac{1}{1-q}} \ \ \ .
\label{eq:ts1}
\end{equation}

It is a~Tsallis--Pareto distribution with the individual energy, $E_i$ and $Z$ is calculated form $\sum p_i=1$.

In high-energy collisions we also have to deal with fluctuations event by event. Following the calculations in references~\cite{artic:tsbentr, artic:tsbphysica}, one may assume that the multiplicity of the created hadrons follow a~negative-binomial distribution for bosons, a~binomial one for fermions. Due to such general reservoir fluctuations, the~$q$ non-extensivity parameter receives a~correction~\cite{artic:tsbentr, artic:tsbphysica}:
\begin{equation}
\label{eq:qwithCandBeta}
q=1-\frac{1}{C}+\frac{\Delta \beta^2}{\left<\beta\right>^2} \ \ \ .
\end{equation}

As the average number of created particles can vary in a~wide range depending on the studied system--typically $\mathcal{O}(10^2)$ in proton-proton collisions, while $\mathcal{O}(10^3-10^5)$ in nucleus-nucleus collisions. We~expect this fluctuation effect to overcome the finite heat capacity condition, therefore one observes $q>1$. It is also straightforward to see that enlarging the system results in $C \to \infty $, and~if fluctuations become sufficiently suppressed, we~get back the Boltzmann--Gibbs case with exponential distribution.
On the other hand, the~assumption $q\rightarrow 1$ leads to a~Gaussian distribution for the $\beta$ values, which can be an~approximation, but never the complete truth. This parameter is used to call the {\it entropic index} or the {\it non-extensivity} parameter, present as a~measure of the deviation from the Boltzmann--Gibbs case, $q=1$. We note, in~high-energy nuclear collisions this value is in the range $1.0<q<1.5$, which~suggests fluctuations override the system size effects, related to the heat capacity of the reservoir.

The~distributions Equations~(\ref{eq:tsexp}) and~(\ref{eq:ts1}) behave similarly, they both can be regarded as Tsallis--Pareto-type distributions. The~authors in \cite{artic:lilin1, artic:lilin2} investigated how the different types fit the experimental data. 
Many further useful readings regarding the thermodynamically consistent non-extensive approach can be found in the literature. The~first possibility is presented in  \cite{artic:wilkentr,artic:actaphys,artic:chaossolitions}, representing the case where the power is proportional to $\frac{1}{q-1}$. An~another kind of approach where the power is $\frac{q}{1-q}$, as discussed in references~\cite{artic:tsthermo1,artic:tsthermo2,artic:tsthermo3,artic:tsthermo4}. 
For our analysis the chosen form is the following~\cite{book:tsb}:
\begin{equation}
\left.\frac{1}{N_{ev}}\frac{\dd^2N}{2\pi p_T \dd p_T \dd y}\right|_{y\approx0}=A\times\left[1+\frac{q-1}{T} (m_T-m) \right]^{-\frac{1}{q-1}} \ \ \
\label{eq:TS}
\end{equation}

As it was shown in \cite{artic:tsbphysica, artic:tsbphysica2} the parameters $q$ and $T$ for an~ideal case are connected to the mean multiplicity and its variance:
\begin{equation}
T=\frac{E}{\left<N\right>}, \ \ \ \ \ q=\frac{\left< N(N+1) \right>}{\left<N \right>^2} \ \ \ .
\label{eq:variance}
\end{equation}

They also may depend on each other. Since in case of $q \to 1$ one has $T \to T_{BG}$, the~parameter $q$ is a~measure of non-extensitivity (i.e.,~non-Gaussivity in $\beta$ fluctuations, non-Poissonity in the multiplicity distribution $P(N)$). $T$ is like the kinetic temperature.

Based on Equation (\ref{eq:variance}), for fixed $\Delta N^2/\left<N\right>^2=\sigma^2$ one obtains:
\begin{equation}
\frac{T}{E}=\sigma^2+(1-q) \ \ \ .
\end{equation}

On the other hand for an~NBD (Negative Binomial Distribution) with fixed $\left<N\right>/k=f$ one gets
\begin{equation}
T=E\times f \times(q-1) \ \ \ .
\end{equation}

Our aim in the followings is to explore the center-of-mass energy evolution of the parameters $q$ and $T$, especially keeping in our mind their physical meaning.
Based on the definition of the PDF and FF of the pQCD-based parton model, we~expect a~logarithmic scaling.
Since this was observed even in fits of electron-positron data~\cite{artic:flowing0}, where PDFs do not appear, we~connect the non-extensive features with the hadronization (fragmentation) processes only.
The~argumentation behind this will be explained in the next subsection.

\subsection{Motivation for Qcd-Like Energy Scaling of the Parameters}
\label{sec:scaling}

Partons, the~elementary momentum carriers in the strongly interacting matter, are tagged with a~quantum number named  \emph{color},
 which property is not observable directly.
The~quarks, antiquarks~and gluons together confine into color singlet hadrons during the hadronization process.
Hadron formation can happen at any energy scale, $Q^2$. The~dependency on it can be factorized into the running nature of the strong interaction's coupling constant, $\alpha(Q^2)$.
Since any observable quantity should be independent of the arbitrary fixing of the energy scale appearing in perturbative QCD calculations, the~mathematical description ought to be (energy-)scale independent at any fixed order. To satisfy this request is not an~easy task, due to the non-perturbative nature of non-Abelian fields at low-energy.

As we presented the hadron production within the pQCD-based parton model in Section~\ref{sec:hadronprod}, the~convolution in Equation (\ref{eq:hadronprod}) includes the scale parameter ($Q^2$) in its kernel.
However, the~cross-section---being an~observable quantity---should be independent of this inner parameter.
Technically, this is achieved with the following method: while calculating the partonic (color) cross sections at a~given order, the~scale can be factorized out and merged into the non-perturbative phenomenological functions.
These are the parton distribution and fragmentation functions, and~they must satisfy a~proper scale evolution equation for avoiding scale-dependent hadronic yields.

To obtain the scale invariance, the~following formula should be fulfilled at any fixed order for any generic form of such phenomenological functions~\cite{artic:dglap1, artic:dglap2, artic:dglap3}:
\begin{equation}
\frac{\partial}{\partial \ln{Q^2}} R(z, Q^2) = 0 \ \ \ ,
\end{equation}
where $R(z, Q^2)$ can be either the PDF or FF, typically at the momentum fraction of the mother and daughter particles, $z$.
In general, this evolution equation determines the possible form of quantity $R$, which naively depends on the current energy scale at a~given fixed order.
Solving this Callan--Symanzik equation one can obtain the energy scale dependence of the running coupling at a~given order in any theory~\cite{artic:pdg}.

In the perturbative QCD based parton model, the~parametrized parton distribution and fragmentation {\em ansatz} functions are typically given in a~power-law form~\cite{artic:ellis}.
One can get then the proper scale evolution by solving the Doksitzer--Gribov--Lipatov--Altarelly--Parisi (DGLAP) equations~\cite{artic:dglap1, artic:dglap2, artic:dglap3}.
Due to its polynomial power-law form, the~predictive power of the calculations gets weaker at low-$z$. We expect that a~Tsallis-like distribution with the appropriate parameter evolution can resolve this problem, providing a~better description.
This motivates us to fit $q$ and $T$ parameters as a~function of the center-of-mass energy $Q^2\sim\sqrt{s}$ in a~similar fashion, as it was done in reference~\cite{artic:gribov80}.
Here our aim is to test the validity of this approach via investigating the energy-evolution dependence of the parameters.


\subsection{The~Improved Quark-Coalescence Model}
\label{sec:coal}

Another description of the hadron formation is based on the constituent quark scaling.
In the \emph{quark coalescence model} the usual underlying assumption is that the hadronization takes place in a~thermal system, where all the participating partons emerge at the same temperature~\cite{artic:tsbeurphys, artic:coalescence1}.
This idea was developed for the description of hadron production in high-energy heavy-ion collisions, where the bulk of the hadron yield closely follows the exponential shape.
In larger colliding systems, like in central collisions of large nuclei, this idea worked well, especially for the low transverse momentum regime, $p_T$ < 3--5 GeV/$c$, with a~single temperature parameter.

\textls[-15]{In the original approach the energy distribution of the partons follow the Boltzmann--Gibbs statistics. Then, one approximates the formation rate as the multiplication of $k$ such Boltzmann--Gibbs distributions:}
\begin{equation}
P_k=\left[f_{BG}(E/k)\right]^k= A' e^{-\beta E} \ \ \ .
\label{eq:Pk}
\end{equation}

In the present non-extensive framework we still assume that the partons are part of a~simple ensemble, but we replace the Boltzmann--Gibbs exponentials by Tsallis--Pareto distributions.
Now the rate is the following:
\begin{equation}
P_k=\left[f_{Ts}(E/k)\right]^k=A^k \left[ 1+ \frac{q-1}{T} E_k \right] ^{-\frac{k}{q-1}} = A' \left[ 1+ \frac{q'-1}{T} E \right] ^{-\frac{1}{q'-1}} \ \ \ ,
\label{eq:Pknonext}
\end{equation}
with $(q'-1)=\frac{(q-1)}{k}$.
In order to test this model, we~check the fitted $q$ parameters of the identified hadrons.
If the equal-energy quark-coalescence theory is correct for both the quark-antiquark containing mesons and triple (anti)quark (anti)baryons, then one observes:
\begin{equation}
(q_{quark}-1)=2(q_{meson}-1)=3(q_{baryon}-1) \ \ \ .
\end{equation}

The~meson-baryon ratio for the Tsallis parameters should be around 3/2:
\begin{equation}
\label{qmperqb}
 \frac{q_{meson}-1}{q_{baryon}-1}=\frac{3}{2} \ .
\end{equation}

This idea formulated by Equation~(\ref{qmperqb}) can be tested as fitting the hadron spectra and investigating the ratio of the parameter $(q_i-1)$ for different identified hadron species. As the constituent quark scaling is getting stronger at larger energies, we~expect to reach this theoretical value only~asymptotically.

\section{Fitted Parameters}
\label{sec:third}

For our analysis we use identified hadron spectra datasets measured in proton-proton collisions in recent years~\cite{artic:62gevphenix, artic:62gevphenix2, artic:200gevphenix, artic:500gevphenix, artic:200gevstar,artic:097tevalice, artic:09tevalice,artic:276tevalice,artic:276tevpi0alice,artic:7tevalice, artic:7tevalice2}.
The~numerical fits of the various datasets were made utilizing the CERN Root analysis software (\myurl{https://root.cern.ch/}, version 6.06/00).
Although we conducted a~comprehensive study, it is worth to note that it is not necessarily meaningful to compare the fit-parameter values for all existing data, because kinematical ranges may vary and the multiplicity-classes are also not defined evenly. This especially holds for kaons and protons at high $p_T$; this generates some uncertainty in those fits. To circumvent these difficulties, we~fixed a~recipe for the fit procedure making it consistently insensitive to the fit program(s), the~size of the fit parameter space and input parameters.

In order to counterbalance the effect of varying ranges we perform the fit procedure in multipl~steps:
\begin{enumerate}[leftmargin=*,labelsep=5mm]
  \item fit of the high-$p_T$ part by fixing $T$ and changing $q$;
  \item fit of the low-$p_T$ part by fixing $q$ and changing $T$;
  \item fit of the whole $p_T$ range with both parameters, starting from the above obtained $q$ and $T$.
\end{enumerate}

We define ``low-$p_T$'' as $p_T< 4$ GeV/$c$ and the ``high-$p_T$'' part as $p_T\geq 2$ GeV/$c$, respectively. The~overlap is intended. In the function defined in Equation~(\ref{eq:TS}) the parameter $T$ sets the characteristic $p_T$ scale. In fact, for $q\longrightarrow 1$ one obtains $f_1(m_T)=Ae^{-(m_T-m)/T}$. The~parameter $q$ on the other hand is linked to the 'power-law like' tail at high $p_T$.
The~procedure was evaluated by comparing the $\chi^2/NDF$~values.

\begin{figure}[H]
\centering
  \centerline{\includegraphics[width=0.98\textwidth]{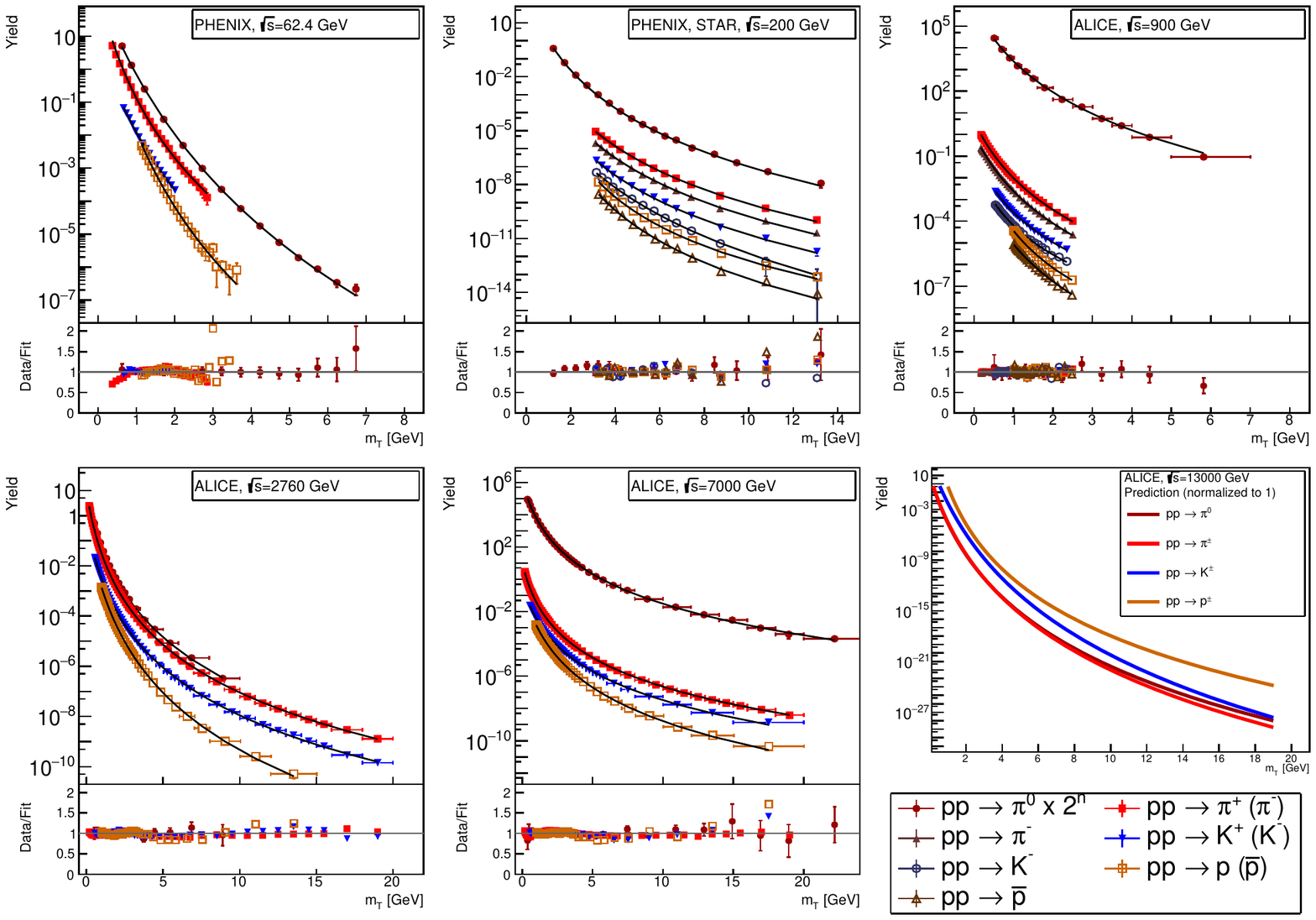}}
  \caption{\textit{Upper panels} are the identified hadron spectra as the function of the transverse mass, $m_T$ is plotted, measured   by the PHENIX~\citep{artic:200gevphenix, artic:62gevphenix, artic:62gevphenix2, artic:500gevphenix}, STAR~\citep{artic:200gevstar}, and~ALICE~\citep{artic:097tevalice,artic:09tevalice,artic:276tevalice,artic:276tevpi0alice,artic:7tevalice,artic:7tevalice2} at different center-of-mass energies from 62.4 GeV $\leq \sqrt{s} \leq 7$ TeV. Experimental data is in comparison with the fitted Tsallis--Pareto functions is indicated as solid lines. The~$m_T$ spectra were scaled by constant factors ($2^m$) for the better visibility as indicated on the graphs. \textit{Lower panels} are present the Data/Fit ratio plots including the estimated fit errors.}
\label{fig:dataperfit}
\end{figure}

The~investigated spectra and the fitted Tsallis--Pareto functions are shown in Figure~\ref{fig:dataperfit}. {\em Upper panels} of the plots present the fits of experimental data measured in proton-proton collisions at $\sqrt{s}=62.4$~GeV, 200 GeV, 500 GeV, 900 GeV, 2.76 TeV, and~7 TeV center-of-mass energies. We considered various neutral, charged and charge-averaged hadron species, $\pi^{\pm}$, $\pi^{0}$, $K^{\pm}$, $p$, and~$\bar{p}$. Identified hadron $m_T$ spectra are scaled by constant factors ($2^n$) for better visibility, as indicated in the panels.

In the {\em lower panels} ``Data/Fit'' plots are presented for each case. One can observe how well the distribution~(\ref{eq:TS}) describes the yields in the whole 62.4 GeV $\leq \sqrt{s} \leq 7$ TeV center-of-mass energy range in the $m_T \lesssim 20$ GeV region. Within the mid $m_T$-regime the overlap with data is excellent, while at the highest $m_T$ values or for heavier hadrons the deviation is somewhat larger.

To determine the center-of-mass energy dependence, we~review the $\sqrt{s}$ evolution of the fitted $q_i$ and $T_i$ parameters for each hadron species, $i \in ~\left\{ \pi^{\pm}, \pi^{0}, K^{\pm}, p, \textrm{and } \bar{p} \right\} $. According to the formula~(\ref{eq:TS}), the~parameters of the Tsallis--Pareto distribution are plotted in Figure~\ref{fig:q,T} as a~function of $\sqrt{s}$. Based on the motivation presented in Section~\ref{sec:scaling}, we~assumed an~energy-evolution for each hadron type $i$ as follows~\cite{artic:karesz}:
\begin{equation}
\label{eq:q-evol}
q_i(\sqrt{s})=q_{1i}+ q_{2i}  \log\left( \sqrt{s}/m_i\right),
\end{equation}
\begin{equation}
\label{eq:T-evol}
T_i(\sqrt{s})=T_{1i}+ T_{2i} \log\left( \sqrt{s}/m_i\right).
\end{equation}

In these formulae, the~mass $m_i$ of the identified hadron $i$ is used to set the physically relevant energy scale.
In Equation (\ref{eq:T-evol}) the parameter $T_{1,i}$ is fixed at $T_{1,i}=50$ MeV, as suggested by reference~\cite{artic:tsbeurphys2}. Hence, the~function in Equation (\ref{eq:T-evol}) is parametrized by $T_{2,i}$.

In Figure~\ref{fig:q,T}a, the~fitted $q_i$ values are plotted for each $\sqrt{s}$, for given identified hadron spectra summarized in Table~\ref{tab:qT}. One can see in the graphs, that the $q_i(\sqrt{s})$ values are close to each other and all curves slightly increase with $\sqrt{s}$, following nicely the formula~(\ref{eq:q-evol}). For pions and kaons the increase is very similar and their evolutions are alike. However, the~precise increase seems to be larger for the~kaons.

We note, that charged and neutral pion results should be consistent. Taking $\pi^0$ and $\pi^{\pm}$ together for the fits, the~mesonic components overlap more. This alternative behavior is thought to be the effect of different kinematical ranges. See more on fitted parameters and $\chi^2/NDF$ in Table~\ref{tab:fitparams} and on kinematical ranges in Table~\ref{tab:kine} in the Appendixes~\ref{appendix:a} and~\ref{appendix:b}.
\begin{table}[H]
\centering
\begin{tabular}{cccccc}
\toprule
\textbf{Hadron,} \boldmath{$i$} & \boldmath{$m_i$} & \boldmath{$q_{1i}$} & \boldmath{$q_{2i}$} & \boldmath{$T_{1i}$} & \boldmath{$T_{2i}$} \\
\midrule
$\pi^{0}$ & 135.0 MeV & $1.03\pm0.002$ & $0.011\pm 0.002 $ & 50 MeV & $0.006\pm 0.001$ MeV \\
$\pi^{\pm}$ & 140.0 MeV & $1.04\pm0.01$ &  $0.009\pm 0.002 $ & 50 MeV & $0.009\pm 0.001 $ MeV  \\
$K^{\pm} $ & 493.0 MeV & $1.00\pm0.01 $ &  $0.016\pm 0.001 $ & 50 MeV & $0.018\pm 0.001 $ MeV  \\
$p(\bar{p})$ & 938.0 MeV & $1.09\pm0.01 $ & $0.004\pm 0.001 $ & 50 MeV & $0.021\pm 0.001 $ MeV \\
\bottomrule
\end{tabular}
\caption{The~$\sqrt{s}$-evolution of the parameters of the fitted Tsallis--Pareto distributions for hadrons, \mbox{$i \in$ $\pi^{\pm}$}, $\pi^{0}$, $K^{\pm}$, $p$, and~$\bar{p}$ in the $62.4$ GeV $\leq \sqrt{s} \leq 7$ TeV c.m. energy range.}
\label{tab:qT}
\end{table}
\unskip
\begin{figure}[H]
    \centering		
		\begin{tabular}{cc}
\includegraphics[width=70mm,scale=0.8,angle=0]{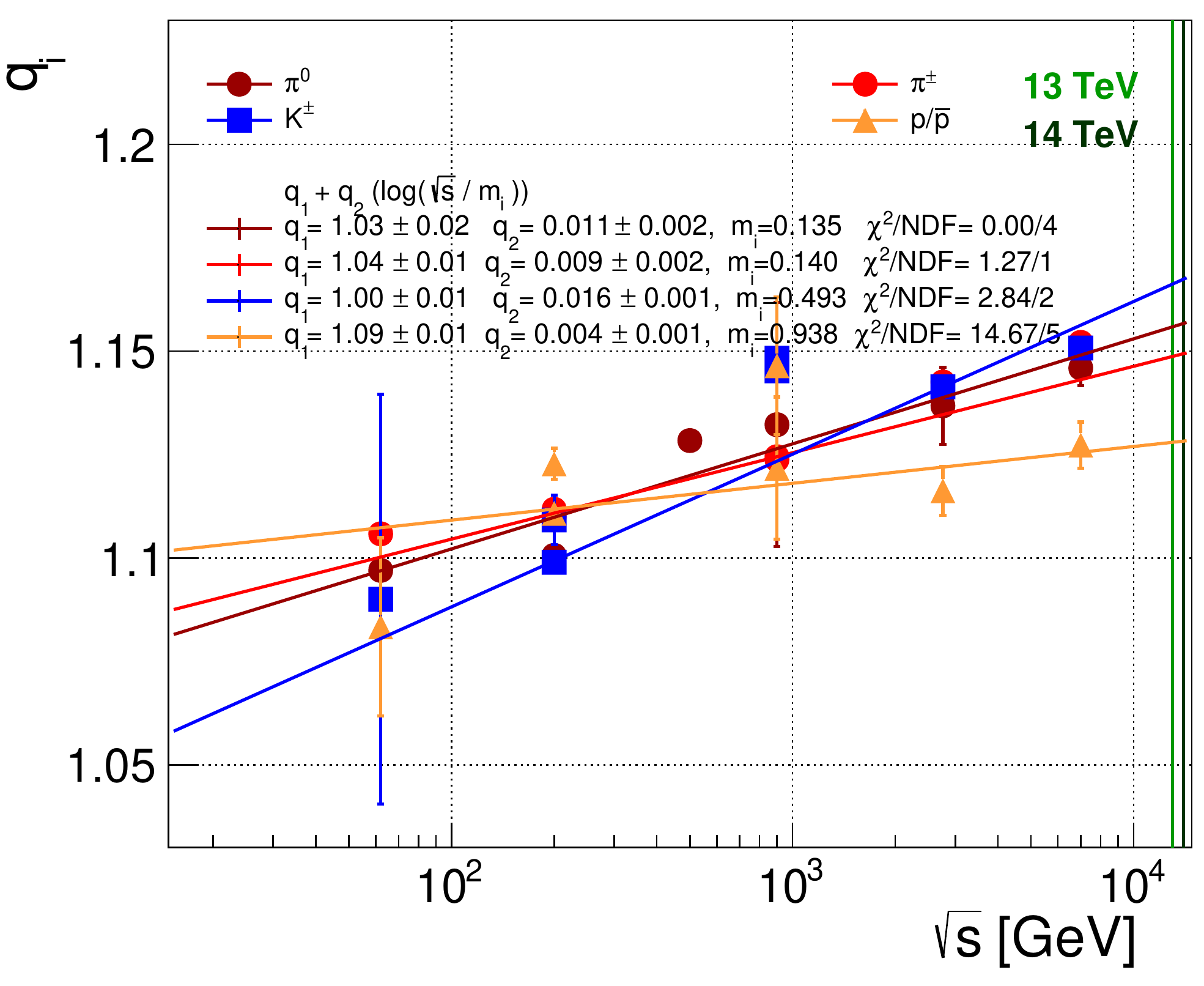}
&     \includegraphics[width=70mm,scale=0.8,angle=0]{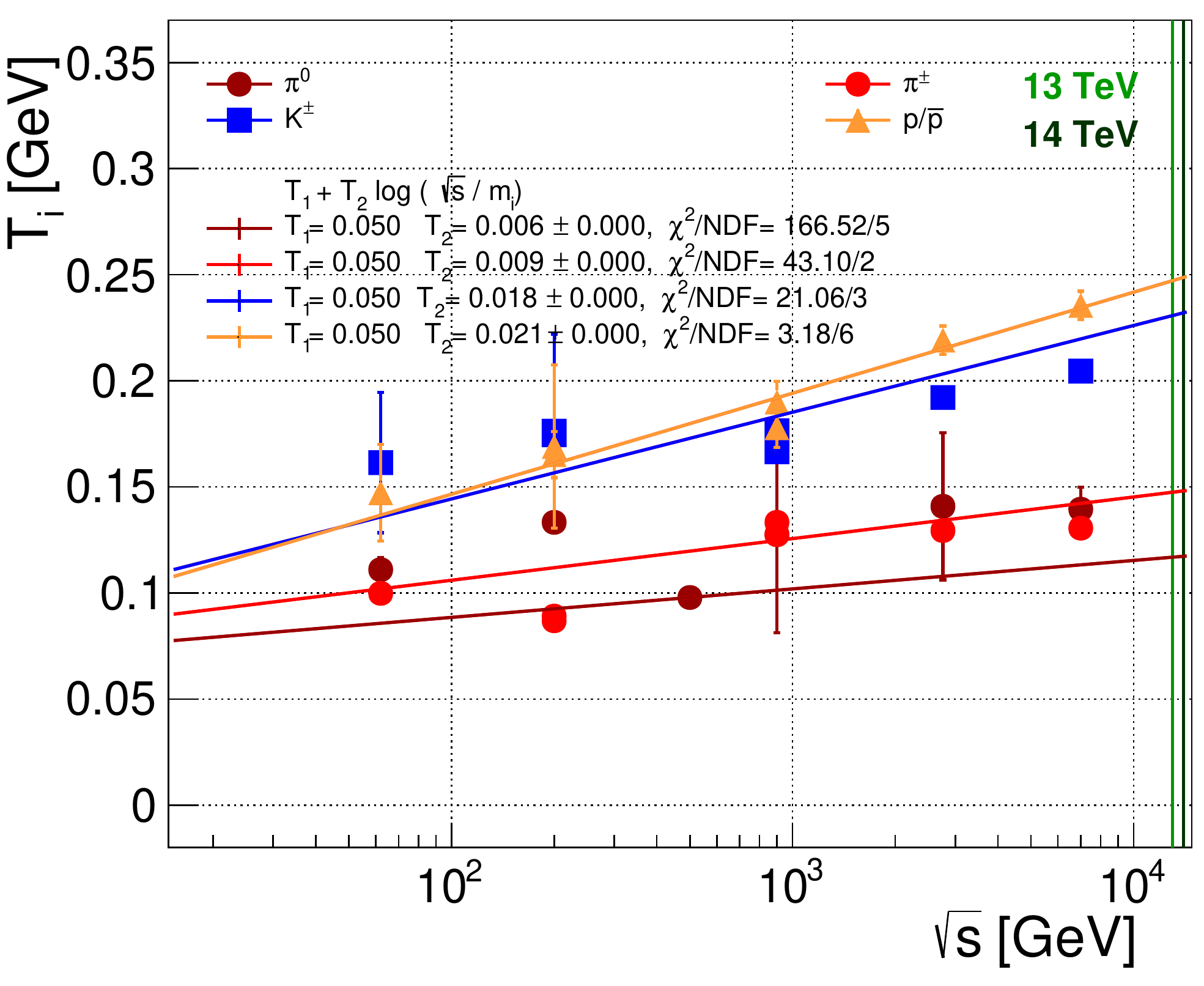}\\
({\bf a})&({\bf b})\\
\end{tabular}

    \caption{The~fitted $q_i$ (\textbf{a}) and $T_i$ (\textbf{b}) as a~function of $\sqrt{s}$ for hadron species, $i$ marked as points. Only the species of the particles are indicated. The~solid color lines are fitted to the pion, kaon, and~(anti)proton points. Vertical lines indicate the place of the $\sqrt{s}=13 $ TeV and $14$ TeV data.}
    		\label{fig:q,T}
\end{figure}

Figure~\ref{fig:q,T}b presents the evolution of the parameter $T_i$ with the center of mass energy. We applied an~evolution according to $\sim \log(\sqrt{s})$ in the Figure~\ref{fig:q,T} using Formula~(\ref{eq:T-evol}) with the evolution parameters listed in Table~\ref{tab:qT}.
Here the energy evolution of the parameter $T(\sqrt{s})$ shows an~increasing trend with the mass of the hadron species. The~obtained parameter values supports the idea of a~mass hierarchy effect: the higher the mass, $m_i$, the~larger $T_{2i}$.

One can realize from Figure~\ref{fig:q,T}a,b, by comparing them in the $\sqrt{s}<3$ TeV regime, that massive protons and kaons present the smaller change in $q_i$, and~their masses are closer to the lattice QCD crossover temperature $T\approx 170$ MeV~\cite{Katz}. Light pions deviate more as increasing the energy, and~the obtained $T_{\pi}$ is smaller, around $\approx 100$ MeV. It is consistent with our picture that, lighter particle can suffer larger fluctuations, which increases the parameter $q_i$ following Equation~(\ref{eq:qwithCandBeta}).

Based on the $\sqrt{s}$ evolution of the experimental fit curves in Figure~\ref{fig:q,T}, we~could predict the parameter values for the soon-to-be available LHC-energy collisions at $\sqrt{s}= 13$ TeV and $14$ TeV. These~energies are indicated on both panels with vertical lines. According to the the assumptions given by the {\em ansatz} Formulae~(\ref{eq:q-evol}) and~(\ref{eq:T-evol}) and the fit parameters from Table~\ref{tab:qT} we summarized these values in Table~\ref{tab:qT-for13TeV} for $\sqrt{s}=13$ TeV. These~data were used to plot the 13 TeV center-of-mass energy prediction on the {\em bottom right panel} of Figure~\ref{fig:dataperfit}. Note, $\sqrt{s}=14$ TeV data is expected to have very similar values within errors.

\begin{figure}[H]
    \centering
		
		\begin{tabular}{cc}
\includegraphics[width=70mm,scale=0.8,angle=0]{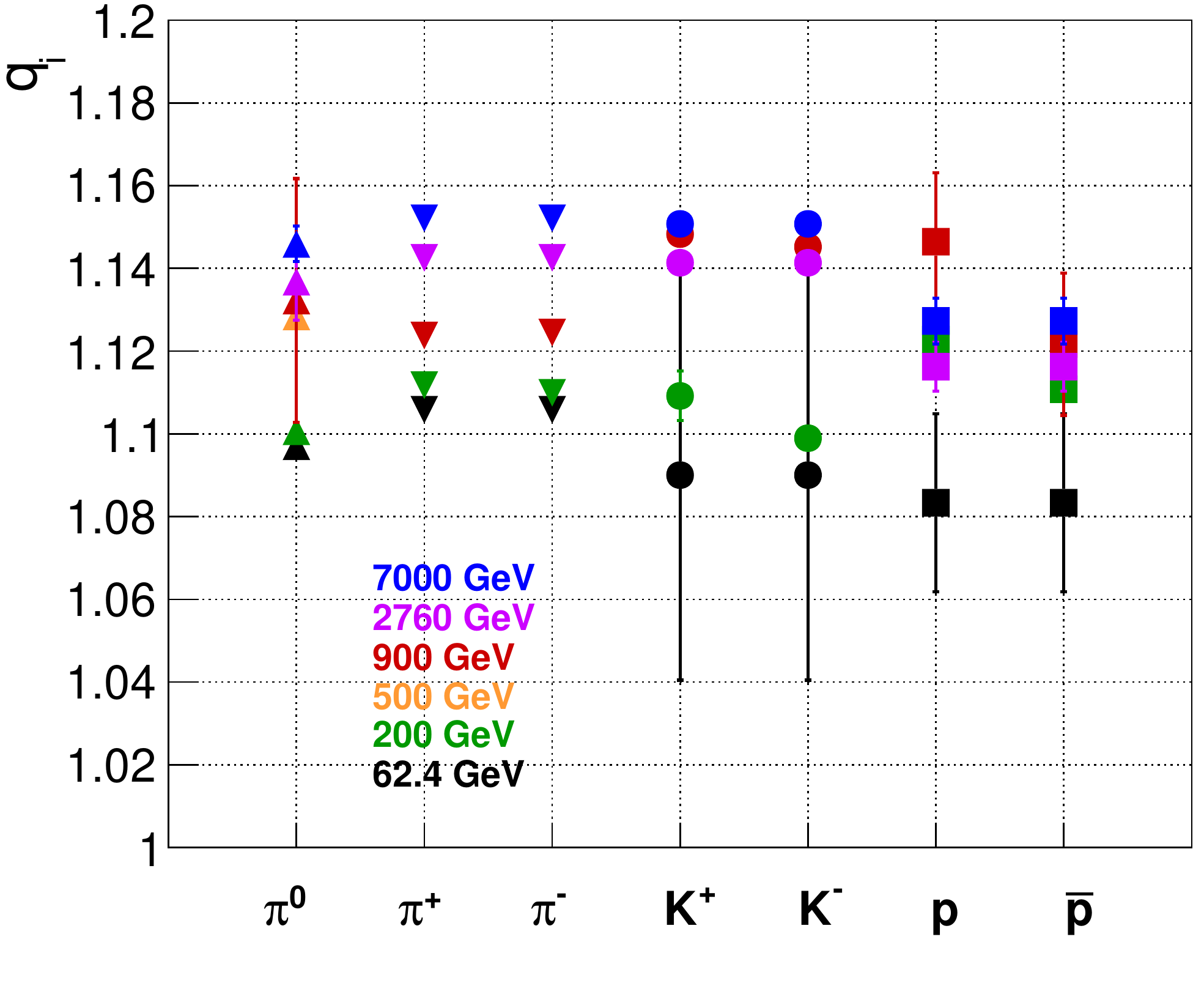}
&     \includegraphics[width=70mm,scale=0.8,angle=0]{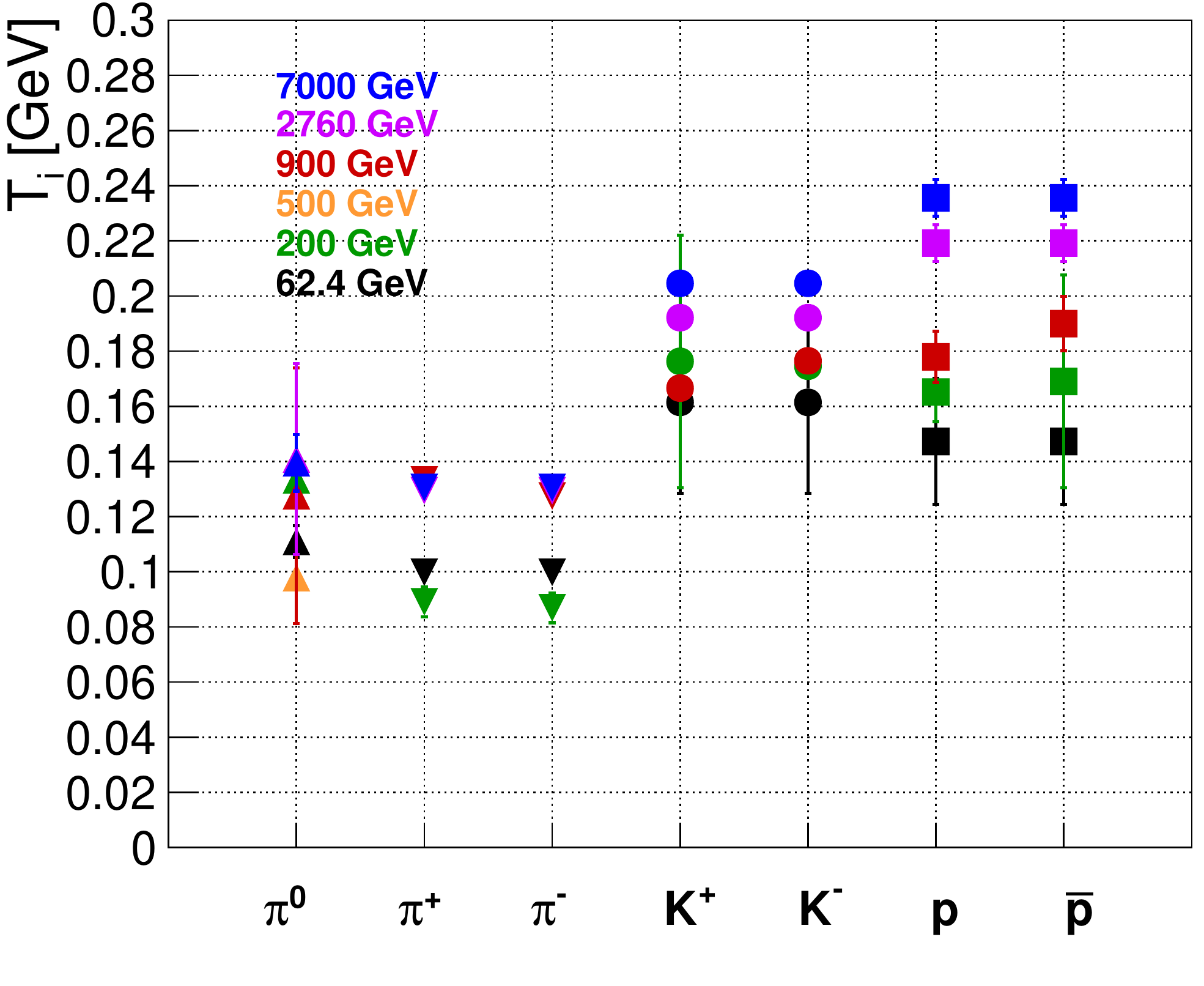}\\
({\bf a})&({\bf b})\\
\end{tabular}

    \caption{The~fitted $q_i$ ({\bf a}) and $T_i$ ({\bf b}) values for each hadron type, $i \in $ $\pi^{\pm}$, $\pi^{0}$, $K^{\pm}$, $K_s^0$, $p$, and~$\bar{p}$, c.f.~\cite{artic:cleymansqspecies}}. 
    \label{fig:qspecies}
\end{figure}

In Figure~\ref{fig:qspecies} we show the fitted $q_i$ and $T_i$ values for different hadron species at the center-of-mass energy values listed above. In agreement with reference~\cite{artic:cleymansqspecies}, we~observe that the non-extensivity parameter $q_i$ on Figure~\ref{fig:qspecies}a is less sensitive to the hadron mass, however the importance of the center-of-mass energy of the colliding system is remarkable. As we have seen  already on Figure~\ref{fig:q,T}, pions have somewhat larger non-extensitivity than more massive hadrons:
\begin{equation}
q_{\pi}(\sqrt{s}) \ \gtrsim \ q_{K}(\sqrt{s}) \ \gtrsim \ q_{p}(\sqrt{s}) \ \ \ .
\label{eq:q-order}
\end{equation}

Figure~\ref{fig:qspecies} also highlights that protons present weaker c.m. energy dependence than mesons.

The~parameter $T_i$ reflects an~opposite mass-hierarchy ordering, on Figure~\ref{fig:qspecies}b. The~more massive hadron, the~larger $T_i$ value:
\begin{equation}
 T_{\pi}(\sqrt{s}) \ < \ T_{K}(\sqrt{s}) \ < \ T_{p}(\sqrt{s})\ \ \ .
\label{eq:T-order}
\end{equation}

Nevertheless, we~observe only a~weak center-of-mass energy dependence.

\begin{table}[H]
\centering
\begin{tabular}{cccc}
\toprule
\textbf{Hadron,} \boldmath{$i$} & \boldmath{$m_i$} & \boldmath{$q_{i}$} & \boldmath{$T_{i}$} \textbf{(MeV)}\\
\midrule
$\pi^{0}$ & 135.0 MeV    & $1.156\pm0.001$/$1.157\pm0.001$ &
$119.0\pm 2.0$/$119.0\pm 2.0$ \\
$\pi^{\pm}$ & 140.0 MeV  & $1.143\pm0.001$/$1.144\pm0.001$ &
$153.0\pm 2.0$/$154.0\pm 2.0$ \\
$K^{\pm} $ & 493.0 MeV   & $1.163\pm0.002$/$1.164\pm0.002$ &
$233.0\pm 2.0$/$234.0\pm 2.0$ \\
$p(\bar{p})$ & 938.0 MeV & $1.128\pm 2.0.003$/$1.128\pm0.003$ &
$250.0\pm 2.0$/$252.0\pm 2.0$ \\
\bottomrule
\end{tabular}
\caption{Predictions for the Tsallis--Pareto parameters for
$\sqrt{s}=13$ TeV ({\it left}) and $14$ TeV ({\it right}), for~hadrons $i \in$ $\pi^{\pm}$, $\pi^{0}$, $K^{\pm}$, $p$, and~$\bar{p}$, based on the Formulae~(\ref{eq:q-evol}) and~(\ref{eq:T-evol}).}
\label{tab:qT-for13TeV}
\end{table}
\unskip

\subsection{The~$(T,q)$ Parameter Space for Identified Hadrons}
\label{sec:qT}

Summarizing our obserations we can conclude that the center-of-mass energy evolution of the fit parameters works well with our logarithmic evolution {\em ansatz}.
\begin{itemize}[leftmargin=*,labelsep=6mm]
\item
The~$q_{2i}$ and $T_{2i}$ parameters are getting slightly larger with the larger hadron mass, $m_i$, and~applying Formulae~(\ref{eq:q-evol}) and~(\ref{eq:T-evol}) the evolution is described nicely in the whole tested energy range, \mbox{$62.4$~GeV $\leq \sqrt{s} \leq 7$ TeV}.

\item The~obtained $q_i(\sqrt{s})$ function increases with $\sqrt{s}$ in the range 1.07--1.17 indicating the deviation from the Boltzmann--Gibbs case where $q=1$. The~deviation from this ``{\em thermodynamical limit}'' case grows as the center-of-mass energy gets higher values. However large statistical errorbars correspond to the lack of statistics in specific particle identification methods of the measurements. (See more in Appendix~\ref{appendix:b}.)

\item The~$T_i(\sqrt{s})$ kinetical temperature parameters almost keep constant values, with the following hadron (mass) hierarchy: $T_{\pi}$ = 120--140 MeV, $T_{K}$ = 120--200 MeV, and~$T_p$ = 70--240 MeV.
\end{itemize}

\begin{figure}[H]
    \centering
		
		\begin{tabular}{cc}
\includegraphics[width=70mm,scale=0.8,angle=0]{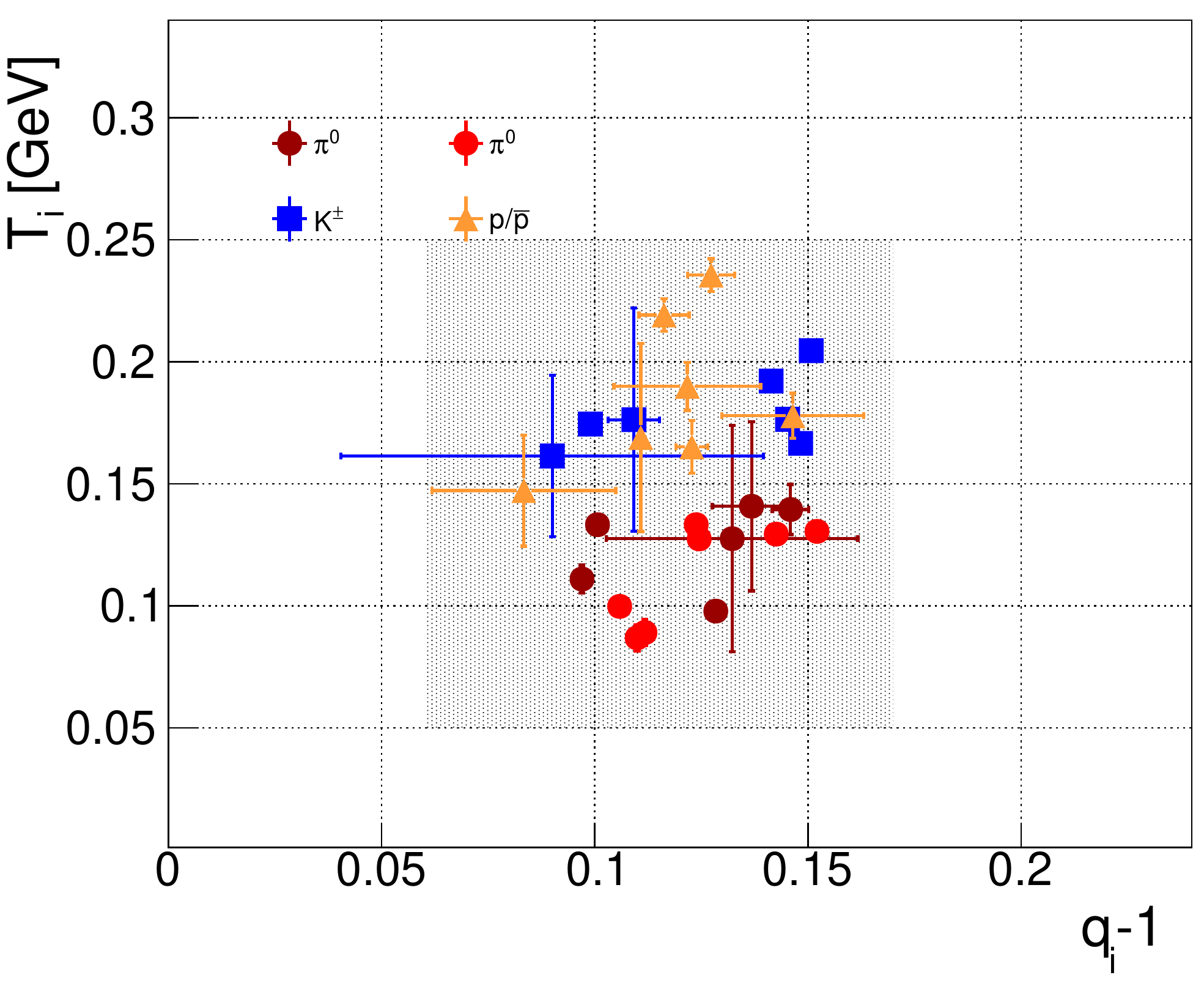}
&     \includegraphics[width=70mm,scale=0.8,angle=0]{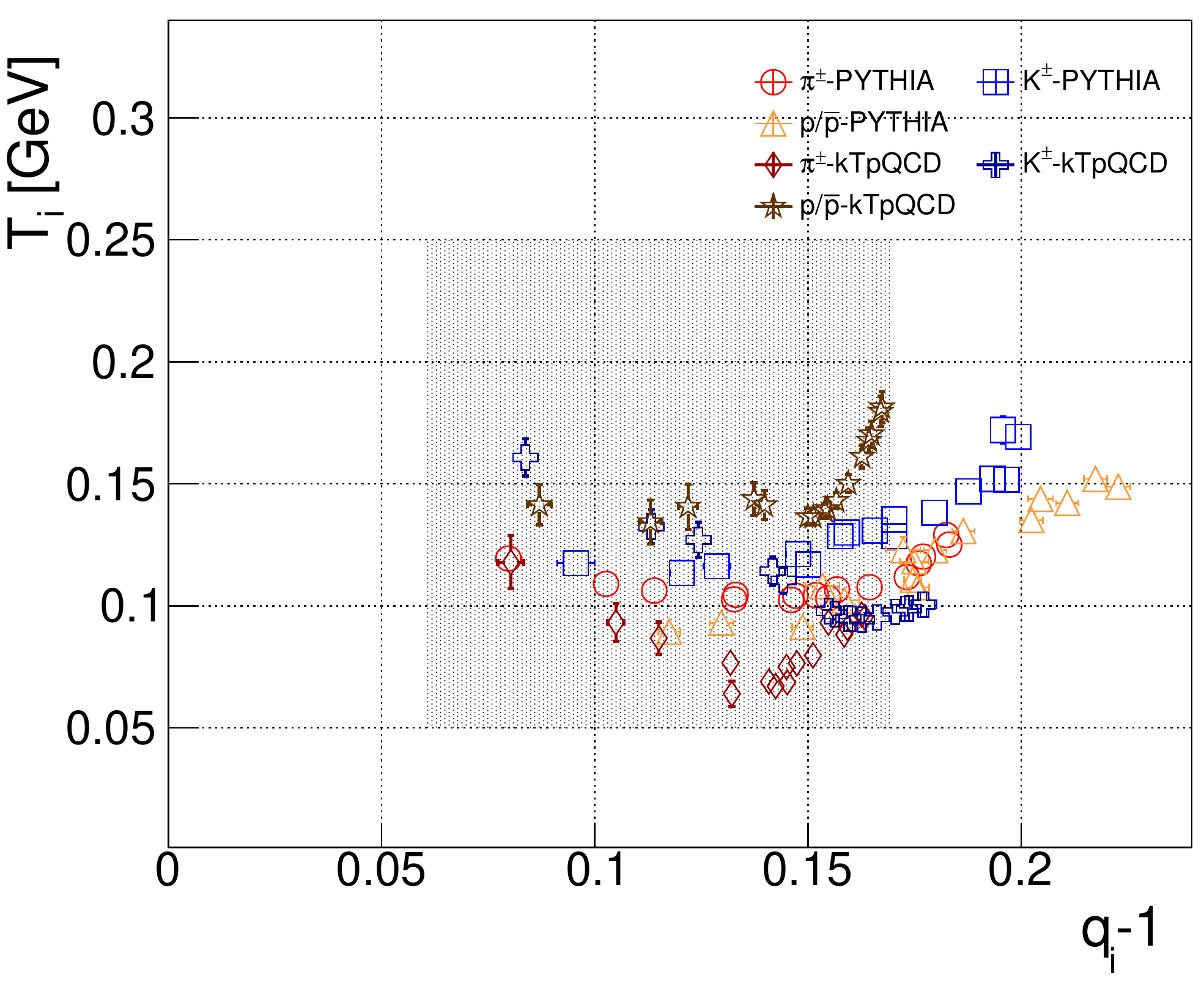}\\
({\bf a})&({\bf b})\\
\end{tabular}

    \caption{(\textbf{a}) The~parameter space $T_i-q_i$ extracted from experimental $p_T$ spectra for hadron types $i \in $ $\pi^{\pm}$, $\pi^{0}$, $K^{\pm}$, $p$, and~$\bar{p}$; (\textbf{b}) The~PYTHIA and kTpQCD\_20 calculated spectra fit results on the same hadron types added to the measurement fit points.}
\label{fig:T-q2}
\end{figure}

We plot the parameters $q_i$ and $T_i$ on the Figure~\ref{fig:T-q2}. The~fitted parameters gather in the $T_{i}\in [70,240]$ MeV and $q_i(\sqrt{s}) \in [1.07,1.17]$ parameter space, which is indicated by the shaded area.

In Figure~\ref{fig:T-q2}b, while keeping the shaded area, we~included the fit-result of theoretical calculations as well. We used two model to get the identified hadron spectra series, namely~PYTHIA8~\cite{artic:pythia6,artic:pythia8} and kTpQCD\_v20~\cite{artic:ktpqcd}. We chose several c.m. energy values and the pseudorapidity region, $|\eta|< 0.5$ for our calculations, and~finally we applied the same fit procedure as described in Section~\ref{sec:fourth}. We plotted the parameters $q_i$ and $T_i$ together with the experimental data-fitted {\em shaded} region.

\vspace{-6pt}

The~calculated points are denoted by {\em empty symbols} for each identified hadron at several center-of-mass energies. Theoretical data fits partially overlap with the {\em shaded} area defined by the experimental fits, although several points are deviating.

\vspace{6pt}
\noindent \textbf{PYTHIA8}
\vspace{6pt}

 We generated 10M events using PYTHIA8~\cite{artic:pythia6,artic:pythia8} Lund's high-energy Monte Carlo event generator to simulate the identified hadron spectra. Points for (charge-averaged) pion, kaon and protons spread in the $q_i \in [1.08,1,23]$ range, wider than the experimental points. In contrast to that, the~$T_i \in [80,150]$ MeV corresponding to the experimental values on the {\em left panel}. Deviating points of the PYTHIA8 results are those which lack sufficient statistics at the highest transverse momenta. Here, the~tail of the distribution is indefinite, thus $q_i$ values fall outside of the experimental $q_i(\sqrt{s}) \in [1.07,1.17]$ parameter space. One recognizes that pions deviate less, since they have the highest statistics among all, followed by kaons and protons. We note that the deviance of $q_i$ disappear as we exclude the low-statistic data at the highest momenta at each energy value, while the consistency with $T_i$ remains.

\vspace{6pt}
\noindent \textbf{kTpQCD\_v20}
\vspace{6pt}

 Hadron spectra calculated within the framework of perturbative QCD were also used utilizing kTpQCD\_v20~\cite{artic:ktpqcd}. These~calculations deliver similar results for all hadron species, because of the similar (polynomial) fragmentation parametrizations. Concerning the correspondence between perturbative QCD results and experimental data fits, both $T_i$ and $q_i$
are running out of the experimental regime in a~similar way. Deviation from the measurement-based data is most remarkable at low c.m. energies, where the $p_T$ range of the spectra is too short due to the limited phase-space. The~domain of pQCD is $p_T > 1.5$ GeV/$c$ and the maximal energy is typically $p_T<\sqrt{s}/2$. This limited range makes the fits more doubtful.

\vspace{6pt}

\begin{figure}[H]
    \centering
		
		\begin{tabular}{cc}
\includegraphics[width=70mm,scale=0.8,angle=0]{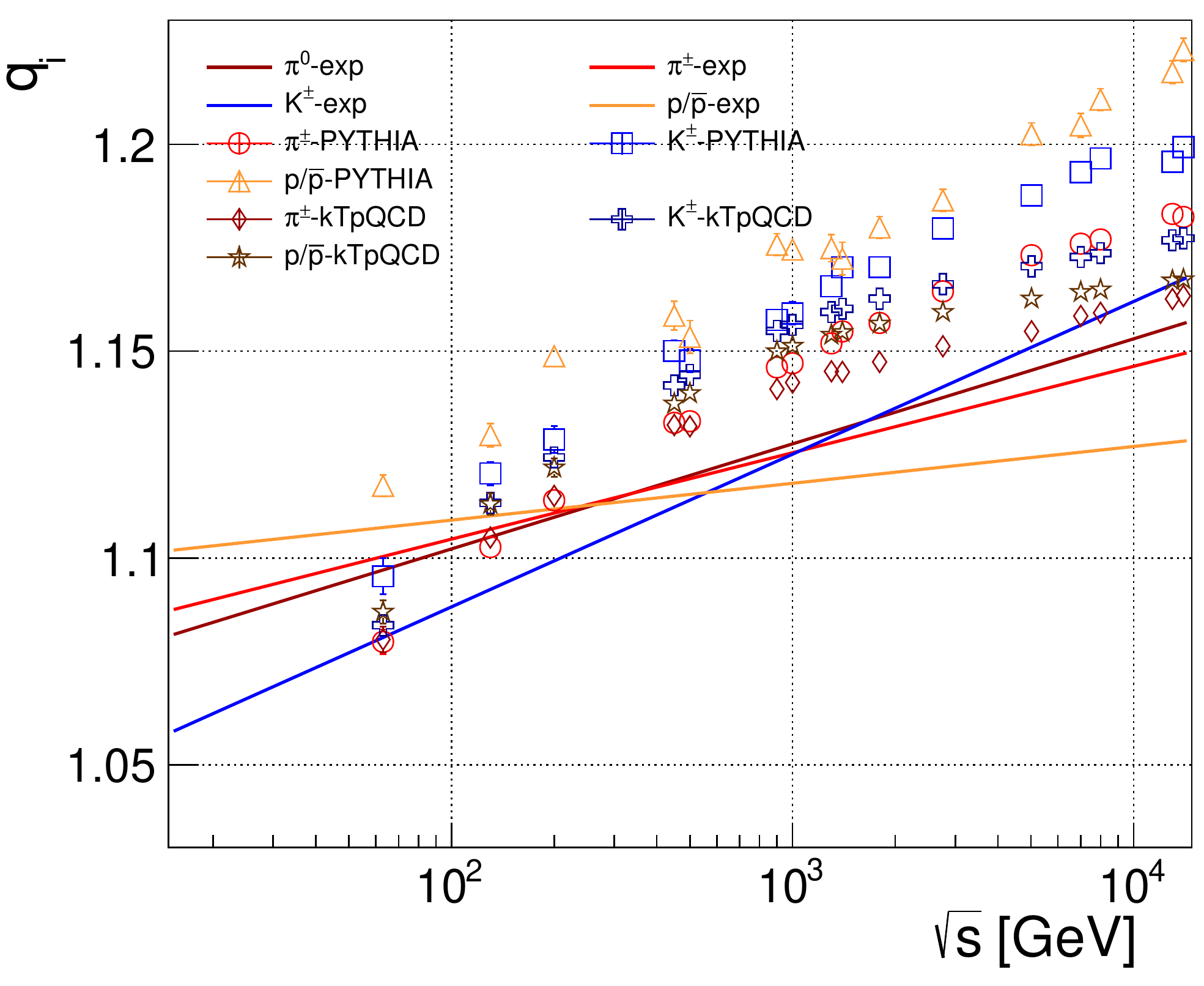}
&     \includegraphics[width=70mm,scale=0.8,angle=0]{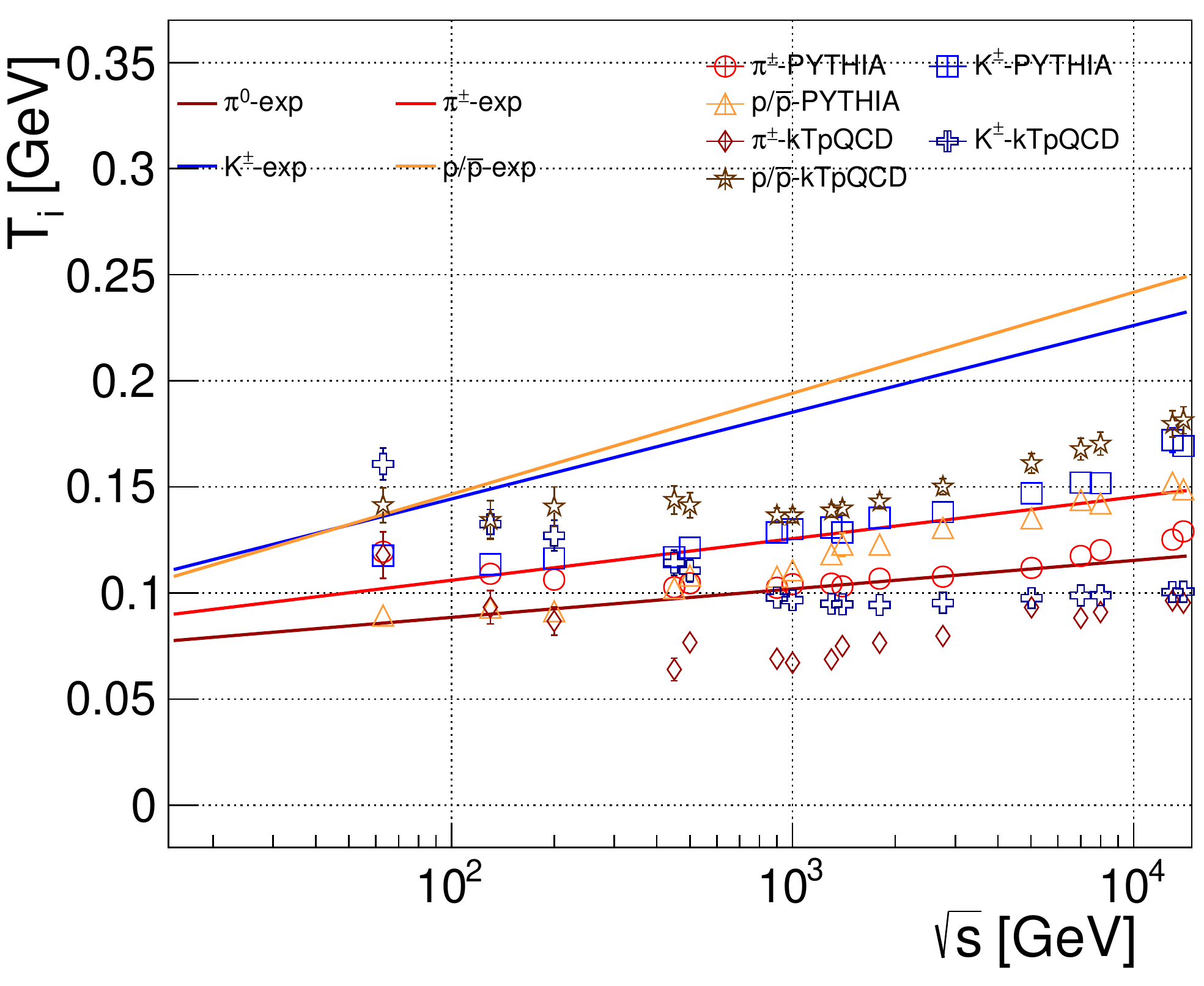}\\
({\bf a})&({\bf b})\\
\end{tabular}

%
    \caption{The~fitted $q_i$ (\textbf{a}) and $T_i$ (\textbf{b}) as a~function of $\sqrt{s}$ for hadron species, $i$ marked as points. Only~the species of the particles are indicated by points. The~solid color lines are fitted to the experimentally measured pion, kaon, and~(anti)proton points.}
\label{fig:qth}
\end{figure}

We also investigated the center-of-mass energy dependence of the fit parameters calculated theoretically from the the PYTHIA8~\cite{artic:pythia6,artic:pythia8} and kTpQCD\_v20~\cite{artic:ktpqcd} models. In Figure~\ref{fig:qth}, we~compare the experimentally observed $\sqrt{s}$-dependence from Figure~\ref{fig:q,T} ({\em solid lines}) to these theoretical model results ({\em data points}). Figure~\ref{fig:qth}a is for parameter $q_i$.

\noindent \textbf{Non-extensivity,} $\boldsymbol{q_i}$:
\vspace{6pt}

According to the above observations, the~perturbative QCD points are close to each other, due to the similar fragmentation function parametrization for the hardon species, mostly at the tail of the distributions at the highest energies and momenta---where experimental and theoretical data meet each other. With kTpQCD\_v20 pions and kaons have similar slopes in $\log(\sqrt{s})$. PYTHIA8 results violate the inequality~(\ref{eq:q-order}) and are larger: $q_{PYTHIA,i} > q_{EXP,i}$. However, the~$\log(\sqrt{s})$ evolution has the same trend and similar slopes---except for protons.

\vspace{6pt}
\noindent \textbf{Temperature-like,} $\boldsymbol{T_i}$:
\vspace{6pt}

The~kTpQCD\_v20 points for each hadron species are close to each other and meet the temperature values only at the highest-energy regime. In this case the formula~(\ref{eq:T-order}) represents a~trend opposite to the perturbative QCD calculations. Theoretical fit parameters deviate here appreciably. We count this for the non-applicability of the pQCD at the low-momentum regime, $p_T<1.5$ GeV/$c$, where the spectra are more thermal-like. On the other hand, PYTHIA8 works well for both the {\em soft} and {\em hard} regimes for the light hadrons. The~$\sqrt{s}$ evolution follows the experimentally observed trend, only~a~small offset is present for kaons and~protons.

We conclude that these model calculations proved their validity in several ways~\cite{artic:predict1, artic:predict2}. Nevertheless, these pictures are not fully consistent with the fit parameters obtained from the data. In~other words, comparison of theoretical models should be made with care within their region of~validity.

\section{Comparison with the Improved Quark-Coalescence Model}
\label{sec:fourth}

As we have explained in Section~\ref{sec:coal}, the~quark-coalescence model was developed for heavy-ion collisions originally, but it was improved by the previously introduced Tsallis distribution. In the followings we endeavor to extend this idea also for smaller systems, such as proton-proton collisions.

\subsection{Connecting Non-Extensivity with the Quark-Coalescence Model}
\label{sec:coal-q}

According to the coalescence picture, the~observed of $(q_{meson}-1)/(q_{baryon}-1)$ should be $3/2$~\cite{artic:quarkscaling}.

\begin{figure}[H]
    \centering
		
		\begin{tabular}{cc}
\includegraphics[width=70mm,scale=0.8,angle=0]{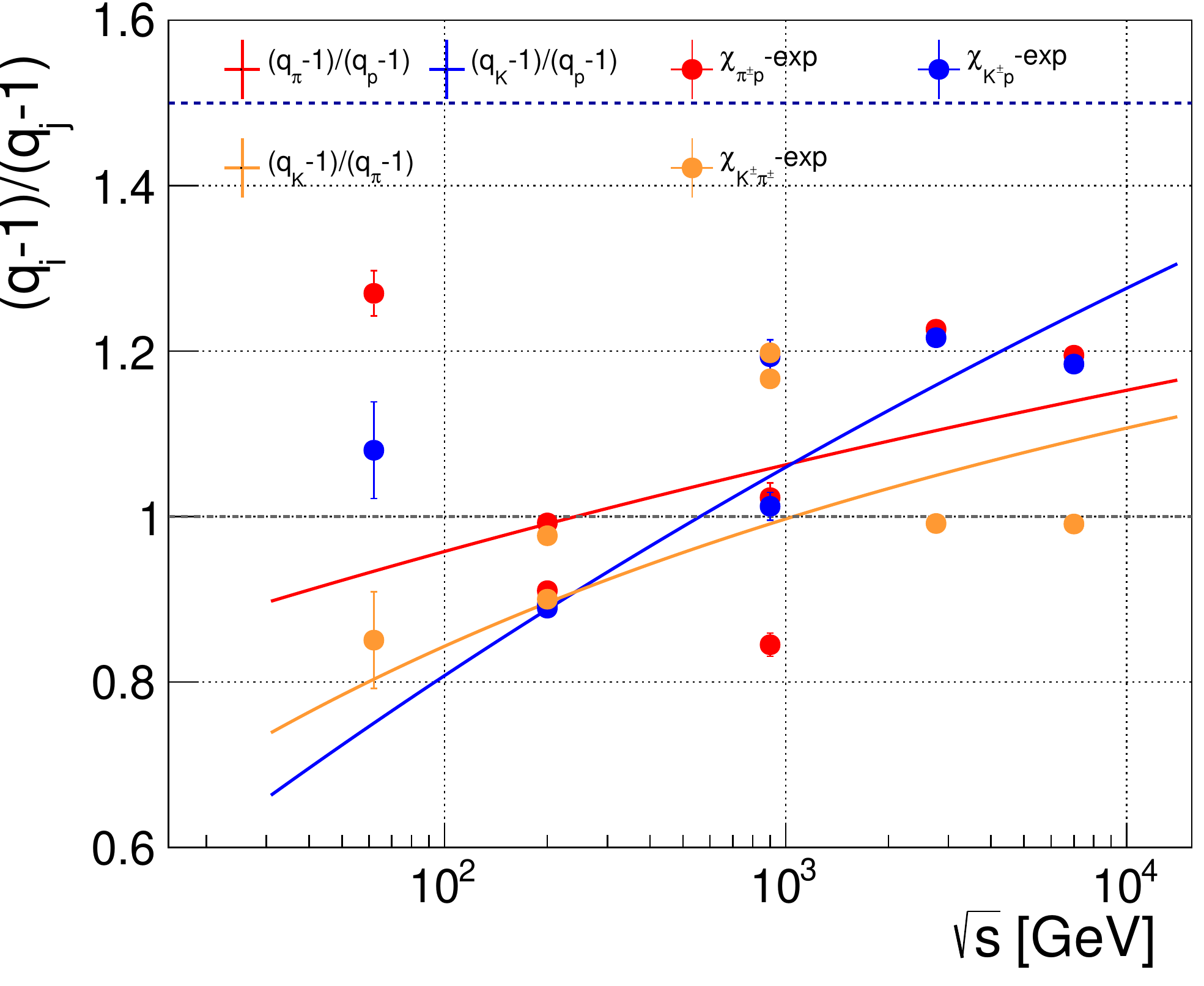}
&     \includegraphics[width=70mm,scale=0.8,angle=0]{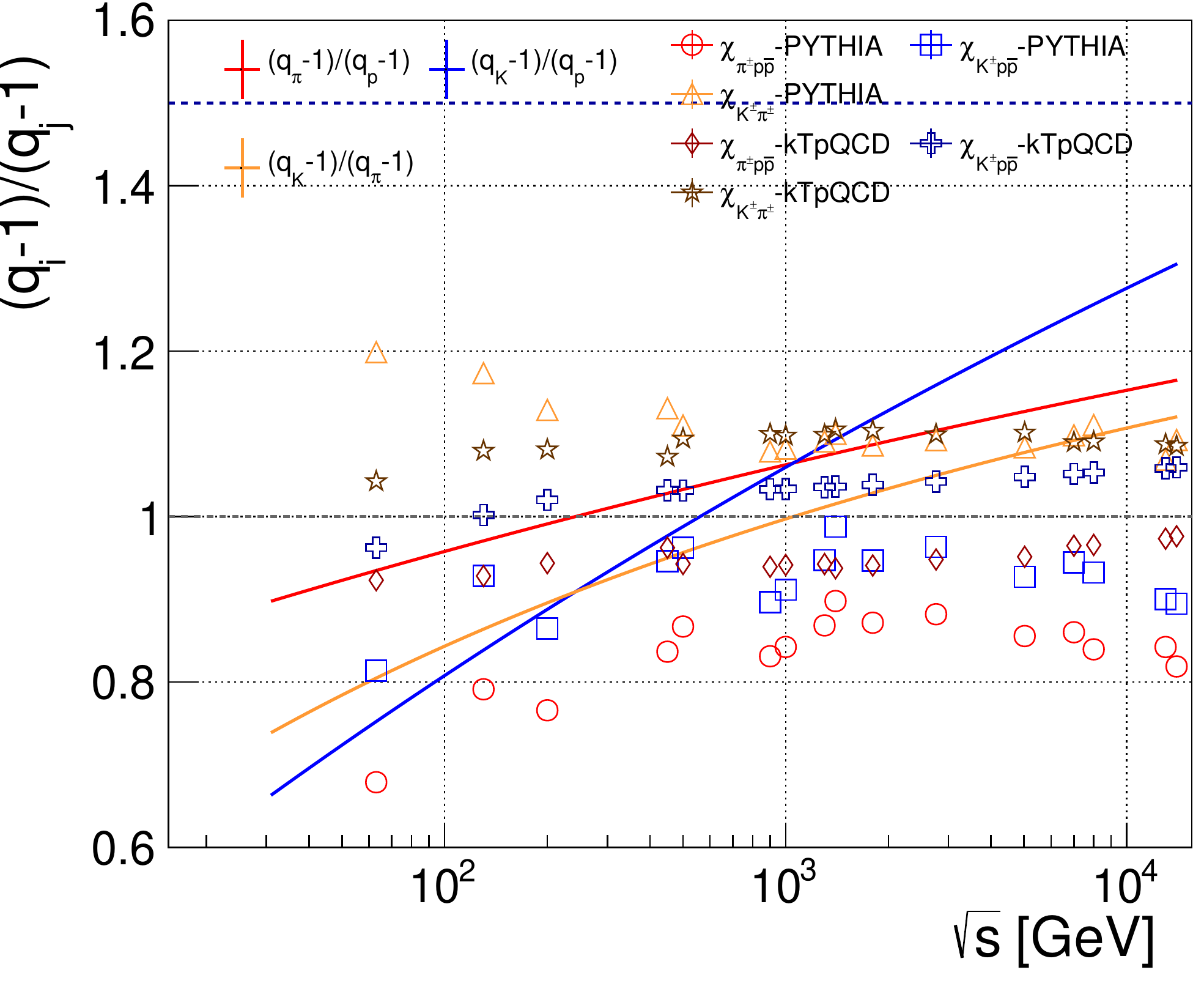}\\
({\bf a})&({\bf b})\\
\end{tabular}
     \caption{ ({\bf a}) Ratio of $(q_{meson}-1) / (q_{baryon}-1)$ plotted in the function of $\sqrt{s}$ as data {\em points} and fit {\em lines}. ({\bf b}) The~PYTHIA and kTpQCD\_20 calculated spectra fit results on the same hadron types added to the measurement fit lines. }
\label{fig:qratios}
\end{figure}

In Figure~\ref{fig:qratios} we plot the ratio $\chi_{ij}=(q_{i}-1)/(q_{j}-1)$ as a~function of the center-of-mass energy, $\sqrt{s}$. Figure~\ref{fig:qratios}a presents the ratios of experimental {\em data points} compared to the fit curve ratios of $\pi/p$, $K/p$ and $K/p$ listed in Table~\ref{tab:qT}. A monotonic, increasing trend of the ratios is clearly seen at the lower energies and saturation can be expected in the most energetic reactions. There are two important observations that is worth note:
\begin{enumerate}[leftmargin=*,labelsep=5mm]
  \item the $(q_{meson}-1)/(q_{baryon}-1)$ fit curves lie below the {\em dashed line} with the value of $3/2$ within the $\sqrt{s} \in [62.4~\textrm{GeV},10~\textrm{TeV}]$ c.m. energy range;
  \item the $\chi_{K\pi}$ kaon-pion ratio shoots over the expected value, 1, a~bit.
\end{enumerate}

On Figure~\ref{fig:qratios}b we plotted the theoretically calculated \emph{points} along with the experimentally fitted {\em solid lines}. It is not surprising, that theoretical curves present quite flat functions for all combination of the ratios, since both the PYTHIA8's Lund fragmentation and the fragmentation function parametrizations in the kTpQCD\_v20 are based on the constituent quark model and the infinite momentum frame is assumed as well. Thus, there is no room for the evolution apart from a~constant hadron-mass effect. This can result only in a~shift of the $q_i$ values as we have seen on the {\em left~panel} of Figure~\ref{fig:qth}. Deviation from the constancy appears only at the lowest energies, but as we have seen earlier in Section~\ref{sec:qT}, at low energies both PYTHIA8 and kTpQCD\_v20 have limited phase space. The~average values of the ratios
$\chi_{ij}=(q_{i}-1)/(q_{j}-1)$ are summarized in Table~\ref{tab:qij} for $i,j \in \{\pi, K, p\}$.
\begin{table}[H]
\caption{The~average values of the hadron spectra parameter ratios, $\chi_{ij}=(q_{i}-1)/(q_{j}-1)$, obtained from theoretical models PYTHIA8 and kTpQCD\_v20.}
\centering
\begin{tabular}{cccc}
\toprule
\textbf{Hadron Ratio} & \textbf{PYTHIA8} & \textbf{kTpQCD\_v20} & \textbf{Colaescence}  \\
\midrule
 $\chi_{K\pi}$ &  $ 1.09 \pm 0.01 $ & $ 1.10 \pm 0.01 $ &  $1.0$ \\
 $\chi_{Kp}$   &  $ 0.95 \pm 0.01 $ & $ 1.06 \pm 0.01 $ &  $1.5$ \\
 $\chi_{\pi p}$&  $ 0.87 \pm 0.01 $ & $ 0.94 \pm 0.01 $ &  $1.5$ \\
\bottomrule
\end{tabular}
\label{tab:qij}
\end{table}

The~kTpQCD\_v20- and PYTHIA8-calculated $(q_{i}-1)/(q_{j}-1)$ points have the same order:
\begin{equation}
\chi_{K\pi}>\chi_{Kp}>\chi_{\pi p}\ \ \ .
\end{equation}

For $\chi_{K\pi}$ both models give the same value, $10\%$ larger than the improved coalescence expectation of 1. For $\chi_{meson,baryon}$ kTpQCD\_v20 has slightly higher values than PYTHIA8, but both are far below the expected value 3/2. Comparing the experimental fit {\em curves} and the theoretically calculated {\em points}, both theory meets the experimental values of $\chi_{K\pi}$. However, for $\chi_{meson,baryon}$ the kTpQCD\_v20 model agrees with the experimental values only at the highest $\sqrt{s}$ c.m. energies.

In summary the improved quark-coalescence model prediction might be reached only beyond the LHC energies, now they seem to support the smaller values. Allegedly, constituent-quark scaling is a~high-$\sqrt{s}$ property. Experimental data support the trends, the~very hadronizaton model needs further~investigation.

\subsection{Investigating the $T_{Slope}$ in the Quark-Coalescence Model}
\label{sec:coal-q}

In the quark-coalescence model, $T_{hadron}=T_{parton}=T$, thus $T_{meson}=T_{baryon}$ is also assumed, c.f.~Section~\ref{sec:coal}. Using the Tsallis distribution we consider the logarithmic slope of $E_i$ spectra:
\begin{equation}
T_{slope}= \left[ -\frac{\dd}{ \dd E_i} \ln P_i \right]^{-1} = T + (q-1)E_i \ \ \ .
\label{eq:tslope}
\end{equation}

\begin{figure}[H]
\centering
  \includegraphics[width=0.75\textwidth]{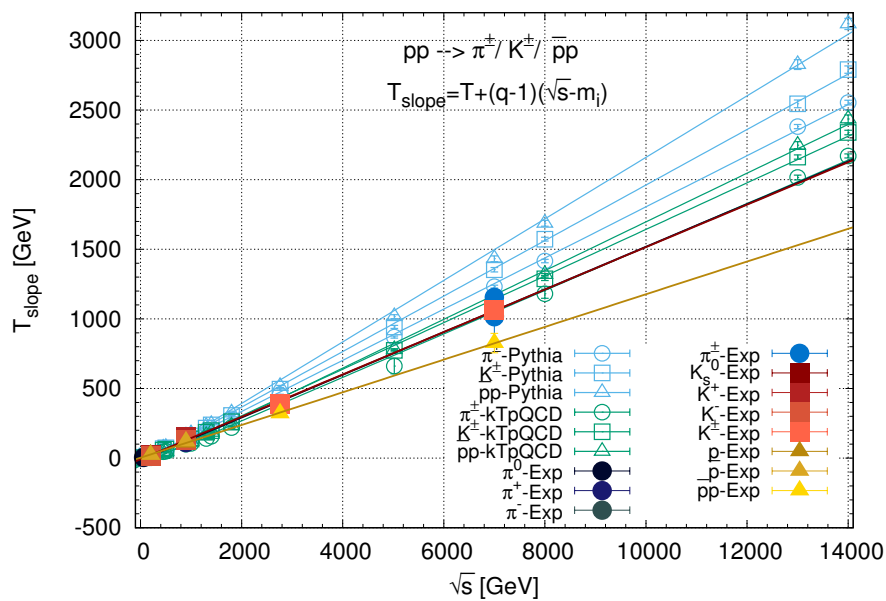}
\caption{The~$T_{slope}(\sqrt{s})$ curves defined by Equation~(\ref{eq:tslope}), fitted on the theoretical PYTHIA8 and kTpQCD\_v20 data (empty points) and on the experimental values (solid points), for all investigated hadron species.}
\label{fig:tslope}
\end{figure}

This may explain the mass ordering found in Equation (\ref{eq:T-order}). A possible way to read off this effect would be to determine the slope of $(m_{T,i}-m_i)$ spectra. Estimating $E_i$ by $\sqrt{s}-m_i$ one obtains results as seen in Figure~\ref{fig:tslope}.

\section{Summary and Discussion}
\label{sec:fifth}

In this study we analyzed identified hadron spectra measured in proton-proton collisions from RHIC to LHC energies in the range $62.4$ GeV $\leq \sqrt{s} \leq 7$ TeV. We showed that the Tsallis--Pareto distributions originated from non-extensive thermodynamics describe the spectra very well in wide $m_T$ regions, typically at $p_T\lesssim$ 10--20 GeV/$c$ using the distribution in the form of Equation~(\ref{eq:TS}).

We provided a~comprehensive and detailed analysis of the state-of-the-art experimental data which will be used also to make predictions about the forthcoming 13 TeV and 14 TeV spectra. The~$\sim\log(\sqrt{s})$-like evolution of the parameters $q_i$ and $T_i$ were tested on the identified hadron spectra data measured for charge averaged $\pi^{\pm}$, $\pi^{0}$, $K^{\pm}$, $p$, and~$\bar{p}$. We observed that both the non-extensivity parameter $q_i$ and temperature-like $T_i$ parameters agree with the suggested QCD-inspired evolution pattern. However, the~temperature has almost a~constant value within the investigated center-of-mass energy regime. We found a~mass-ordered hierarchy in the evolution parameters of the experimental fits, i.e.,~lighter hadron spectra have the more non-extensive $q_{i}>1$ and heavier hadron spectra fit with larger $T_i$ values. However, we~note that the deduced $T_i$ values might be sensitive not just the hadron mass but also on the chosen distribution function. This might result different values extracted from similar data, e.g.,~in reference~\cite{artic:cleymansTS}.

We compared the experimental data fit results with theoretical predictions. The~c.m. energy evolution of the fit parameters were calculated by PYTHIA8~\cite{artic:pythia6,artic:pythia8} and kTpQCD\_v20~\cite{artic:ktpqcd}.
We~found theoretical fit parameters to be more compact in the $(T,q)$ space than the experimental ones: \mbox{$T_i \in [80,240]$ MeV,} while non-extensivity is wider than the measurement-based $q_i \in [1.08,1,23]$ region. The~most deviating points arise from limitations of the theoretical models, i.e.,~where statistics is low or the phase-space is limited. Energy evolution in the theoretical models were investigated as well. In~agreement with our expectations we conclude that
\begin{enumerate}[leftmargin=21pt,labelsep=7pt]
  \item[(i)] for the $\sqrt{s}$ evolution, kTpQCD\_v20 agrees more with the power-law related non-extensivity parameter $q_i$;
  \item[(ii)] PYTHIA8 results correspond well with the measured $T_i(\sqrt{s})$ evolution.
\end{enumerate}

The~study of these models reflected the lack of the proper handling of the hadron-mass, since all assumptions fail for more massive hadron species.

At the highest energies and momenta, in~the infinite momentum frame, constituent quark number scaling is assumed to get stronger. To test this idea in the framework of the non-extensive approach, we~applied and investigated an~improved quark-coalescence model, inserting Tsallis-like energy distribution kernels where constituent quark scaling appears explicitly. Experimental~data present a~slight monotonic-increase with c.m. energy, but the saturation ridge of the ratio \mbox{$(q_{meson}-1)/(q_{baryon}-1)$} is lower than predicted by the coalescence-theory (3/2) while the reference ratio $(q_{meson}-1)/(q_{meson'}-1)$ is only slightly apart from the expected value $1$.
The~fit parameters calculated by PYTHIA8 and kTpQCD\_v20 models both have almost no $\sqrt{s}$ evolution, but only the ratio values for light mesons are in agreement with the experimental data especially at the highest LHC energies.

In summary, our detailed analysis aimed to investigate how we can provide physical meaning for experimentally-fitted parameters, based on well-known theoretical models and phenomena. Our~results motivate us to improve the model of hadronization in high-energy collisions, using spectra with exponential shape at low-$p_T$, keeps the power-law tail at thigh $p_T$, and~takes care of the meson/baryon spectra ratios and/or the experimentally observed $(q_{meson}-1)/(q_{barion}-1)$.

\vspace{6pt}


\acknowledgments{This work has been supported by Hungarian OTKA grants K104260, NK106119, K120660, by~the MTA-UA bilateral mobility program NKM-81/2016, and~by the Hungarian-Chinese Collaboration NKFIH TET 12 CN-1-2012-0016. G.G.B. thanks the J\'anos Bolyai Research Scholarship of the Hungarian Academy of Sciences for support. G.B. thanks for the support of Wigner GPU Laboratory. Author \'A.T. is supported by the \'UNKP-16-2 New National Excellence Program of the Ministry of Human Capacities, Hungary.}

\authorcontributions{G.B. collected and elaborated the experimental data and wrote the manuscript. G.G.B. conceived the study, interpreted the data and wrote the first version of the paper. \'A.T. performed the theoretical model calculations and analyzed the simulated data. T.S.B. developed the theoretical background and improved the formulations of the manuscript. K.\"U. initiated the method used for interpretation of the data presented in this~work.}

\conflictsofinterest{The~authors declare no conflict of interest.}


\abbreviations{The~following abbreviations are used in this manuscript:\\

\noindent 
\begin{tabular}{@{}ll}
ALICE & A Large Ion Colliding Experiment \\
BNL & Brookhave National Laboratory \\
CERN & Conseil Européen pour la Recherche Nucléaire \\
CM & Center-of-mass \\
DGLAP & Dokshitzer--Gribov--Lipatov--Altarelli--Parisi \\
FF & Fragmentation Function\\

\end{tabular}} 

\noindent 
\begin{tabular}{@{}ll}
LHC & Large Hadron Collider \\
NBD & Negative Binomial Distribution \\
NDF & Number of Degrees of Freedom \\
PHENIX & A Physics Experiment at RHIC \\
PDF & Parton Distribution Function\\
RHIC & Relativistic Heavy Ion Collider \\
(p)QCD & (perturbative) Quantum Chromo Dynamics \\
STAR & Solenoidal Tracker at RHIC \\
QGP & Quark Gluon Plasma
\end{tabular}
\appendixtitles{no} 
\appendixsections{multiple} 
\appendix

\setcounter{table}{0}
\renewcommand{\thetable}{A\arabic{table}}

\section{}
\label{appendix:a}
In Table \ref{tab:fitparams} we show the parameters fitted to all used datasets~\cite{artic:62gevphenix,artic:62gevphenix2,artic:200gevphenix,artic:200gevstar, artic:500gevphenix,artic:09tevalice,artic:097tevalice,artic:276tevalice,artic:276tevpi0alice,artic:7tevalice,artic:7tevalice2}.
The~parameters $q_i$, $T_i$ and $A_i$ from Equation (\ref{eq:TS}). and the $\chi^2/$ndf values of the given fit are listed for each identified hadron and for each $\sqrt{s}$ center-of-mass energy value.

\begin{table}[H]
\caption{The~fitted $q$, $T$ and $A$ parameters and the $\chi^2$/ndf value of the fit for identified hadrons $\pi^0$, $\pi^\pm$, $K^\pm$ and $p\bar{p}$. The~values are grouped according to the hadrons such that the change in respect to the center-of-mass energy can be compared easily.}
\centering
\small
\scalebox{0.93}[0.93]{\begin{tabular}{ccccccc}
\toprule
\boldmath{$\sqrt{s}$} \textbf{(GeV)} &  \textbf{Hadron} & \boldmath{$q$} & \boldmath{$T$} \textbf{(GeV)} & \boldmath{$A$} & \boldmath{$\chi^2/$}\textbf{ndf} & \textbf{Experiment} \\
\midrule
62	  &$\pi^{0}$   & 1.073 $\pm$ 0.011 & 0.139 $\pm$ 0.027 & 121.252 $\pm$ 100.460       & 7.857/11    & PHENIX~\cite{artic:62gevphenix2}\\
200	  &$\pi^{0}$   & 1.101 $\pm$ 0.002 & 0.133 $\pm$ 0.003 & 149.459 $\pm$ 15.459        & 6.450/14    & PHENIX~\cite{artic:62gevphenix} \\
500	  &$\pi^{0}$   & 1.128 $\pm$ 0.000 & 0.098 $\pm$ 0.000 & 746.070 $\pm$ 0.527         & 3656.733/25 & PHENIX~\cite{artic:500gevphenix}\\
900	  &$\pi^{0}$   & 1.132 $\pm$ 0.029 & 0.128 $\pm$ 0.046 & 302183.366 $\pm$ 321455.697 & 0.462/10    & ALICE~\cite{artic:097tevalice}\\
2760	&$\pi^{0}$   & 1.137 $\pm$ 0.009 & 0.141 $\pm$ 0.035 & 4.937 $\pm$ 5.116           & 0.237/15    & ALICE~\cite{artic:276tevpi0alice}\\
7000	&$\pi^{0}$   & 1.146 $\pm$ 0.004 & 0.140 $\pm$ 0.010 & 498950.603 $\pm$ 133429.129 & 1.143/30    & ALICE~\cite{artic:097tevalice}\\
62	  &$\pi^{\pm}$ & 1.106 $\pm$ 0.000 & 0.100 $\pm$ 0.000 & 245.248 $\pm$ 0.000         & 4.276/23    & PHENIX~\cite{artic:62gevphenix}\\
200	  &$\pi^{+}$   & 1.112 $\pm$ 0.001 & 0.089 $\pm$ 0.005 & 39.389 $\pm$ 14.396         & 58.248/11   & STAR~\cite{artic:200gevstar}\\
200	  &$\pi^{-}$   & 1.110 $\pm$ 0.001 & 0.087 $\pm$ 0.005 & 48.554 $\pm$ 18.250         & 75.053/11   & STAR~\cite{artic:200gevstar}\\
900	  &$\pi^{+}$   & 1.124 $\pm$ 0.000 & 0.133 $\pm$ 0.000 & 4.496 $\pm$ 0.000           & 4.413/12    & ALICE~\cite{artic:09tevalice}\\
900	  &$\pi^{-}$   & 1.124 $\pm$ 0.002 & 0.127 $\pm$ 0.001 & 5.240 $\pm$ 0.082           & 10.816/30   & ALICE~\cite{artic:09tevalice}\\
2760	&$\pi^{\pm}$ & 1.143 $\pm$ 0.000 & 0.129 $\pm$ 0.000 & 12.546 $\pm$ 0.000          & 3.929/60    & ALICE~\cite{artic:276tevalice}\\
7000	&$\pi^{\pm}$ & 1.152 $\pm$ 0.000 & 0.131 $\pm$ 0.000 & 14.544 $\pm$ 0.000          & 5.750/55    & ALICE~\cite{artic:7tevalice, artic:7tevalice2}\\
62	  &$K^{\pm}$   & 1.090 $\pm$ 0.050 & 0.161 $\pm$ 0.033 & 3.142 $\pm$ 1.155           & 0.192/13    & PHENIX~\cite{artic:62gevphenix}\\
200	  &$K^{+}$     & 1.109 $\pm$ 0.001 & 0.122 $\pm$ 0.002 & 0.902 $\pm$ 0.187           & 31.664/12   & STAR~\cite{artic:200gevstar}\\
200	  &$K^{-}$     & 1.083 $\pm$ 0.005 & 0.199 $\pm$ 0.116 & 0.091 $\pm$ 0.058           & 17.189/11   & STAR~\cite{artic:200gevstar}\\
900	  &$K^{+}$     & 1.148 $\pm$ 0.000 & 0.167 $\pm$ 0.000 & 0.203 $\pm$ 0.000           & 6.932/24    & ALICE~\cite{artic:09tevalice}\\
900	  &$K^{-}$     & 1.145 $\pm$ 0.000 & 0.176 $\pm$ 0.000 & 0.186 $\pm$ 0.000           & 19.465/24   & ALICE~\cite{artic:09tevalice}\\
2760	&$K^{\pm}$   & 1.141 $\pm$ 0.002 & 0.192 $\pm$ 0.004 & 0.434 $\pm$ 0.022           & 2.793/55    & ALICE~\cite{artic:276tevalice}\\
7000	&$K^{\pm}$   & 1.151 $\pm$ 0.000 & 0.205 $\pm$ 0.000 & 0.500 $\pm$ 0.000           & 3.756/48    & ALICE~\cite{artic:7tevalice, artic:7tevalice2}\\
62	  &$p/\bar{p}$ & 1.083 $\pm$ 0.022 & 0.147 $\pm$ 0.023 & 1.240 $\pm$ 0.440           & 4.722/24    & PHENIX~\cite{artic:62gevphenix}\\
200	  &$p$         & 1.118 $\pm$ 0.001 & 0.070 $\pm$ 0.001 & 10.205 $\pm$ 13.089         & 15.983/11   & STAR~\cite{artic:200gevstar}\\
200	  &$\bar{p}$   & 1.109 $\pm$ 0.001 & 0.074 $\pm$ 0.003 & 9.945 $\pm$ 2.227           & 26.393/11   & STAR~\cite{artic:200gevstar}\\
900	  &$p$         & 1.146 $\pm$ 0.017 & 0.178 $\pm$ 0.009 & 0.053 $\pm$ 0.003           & 13.758/21   & ALICE~\cite{artic:09tevalice}\\
900	  &$\bar{p}$   & 1.122 $\pm$ 0.017 & 0.190 $\pm$ 0.010 & 0.049 $\pm$ 0.002           & 13.337/21   & ALICE~\cite{artic:09tevalice}\\
2760	&$p/\bar{p}$ & 1.116 $\pm$ 0.006 & 0.219 $\pm$ 0.007 & 0.110 $\pm$ 0.005           & 2.232/45    & ALICE~\cite{artic:276tevalice}\\
7000	&$p/\bar{p}$ & 1.127 $\pm$ 0.006 & 0.236 $\pm$ 0.007 & 0.117 $\pm$ 0.005           & 2.556/46    & ALICE~\cite{artic:7tevalice, artic:7tevalice2}\\
\bottomrule
\end{tabular}}

\label{tab:fitparams}
\end{table}

\section{}
\label{appendix:b}
In Table \ref{tab:kine} the used datasets, their kinematical properties and the corresponding references are listed.
As we already mentioned in Section \ref{sec:third}. that the spectra were measured in wide range of kinematical variables, although these values varies in different experiments.

\begin{table}[H]
\begin{center}
\small
\begin{tabular}{ccccc}
\toprule
\boldmath{$\sqrt{s}$} \textbf{(TeV)} & \textbf{Rapidity} & \textbf{Particle} & \boldmath{$p_T$} \textbf{Range [GeV/}\boldmath{$c$}\textbf{]} & \textbf{Experiment} \\
\midrule
0.062 & $|\eta|<0.35$& $\pi^0$ & $0.5 \leq p_T \leq7$  & PHENIX~\cite{artic:62gevphenix2} \\
  & &                 $\pi^{\pm}$    & $0.3 \leq p_T \leq 3$  & PHENIX~\cite{artic:62gevphenix}\\
  & &                 $K^\pm$     & $0.5 \leq p_T \leq 2$  & \\
  & &                 $p/\bar{p}$ & $1 \leq p_T \leq 4$ & \\
\midrule
0.2    & $|\eta|<0.35$ & $\pi^0$   & $1 \leq p_T \leq 15$ & PHENIX~\cite{artic:200gevphenix} \\
  &       $|y|<0.5$     & $\pi^\pm$ & $3 \leq p_T\leq15$  & STAR~\cite{artic:200gevstar} \\
  & &                     $K^\pm$    & $3 \leq p_T\leq15$  & \\
  & &                   $p/\bar{p}$ & $3 \leq p_T\leq15$ & \\
\midrule
0.5    & $|\eta|<0.35$ & $\pi^0$ & $1 \leq p_T \leq 30$  & PHENIX~\cite{artic:500gevphenix} \\
\midrule
0.9    & $|y|<0.5$  &     $\pi^0$     & $0.4 \leq p_T \leq 7$ & ALICE~\cite{artic:097tevalice} \\
  & &                   $\pi^\pm$   & $0.1 \leq p_T \leq 2.5$ & ALICE~\cite{artic:09tevalice} \\
  & &                   $K^\pm$     & $0.2 \leq p_T \leq 2.5$ & \\
  & &                   $p/\bar{p}$ & $0.3 \leq p_T \leq 2.5$ & \\
\midrule
2.76   & $|\eta|<0.8$  &$\pi^0$     & $0.6 \leq p_T \leq 20$ & ALICE~\cite{artic:276tevpi0alice} \\
  & &                    $\pi^\pm$   & $0.1 \leq p_T \leq 20$ & ALICE~\cite{artic:276tevalice} \\
  & &                    $K^\pm$     & $0.1 \leq p_T \leq 20$ & \\
  & &                    $p/\bar{p}$ & $0.3 \leq p_T \leq 15$ & \\
\midrule
7      & $|y|<0.5$  & $\pi^0$     & $0.3 \leq p_T \leq 25$ & ALICE~\cite{artic:097tevalice} \\
  & &                  $\pi^\pm$   & $0.1 \leq p_T \leq 20$ & ALICE~\cite{artic:7tevalice, artic:7tevalice2} \\
  & &                  $K^\pm$     & $0.2 \leq p_T \leq 20$ & \\
  & &                  $p/\bar{p}$ & $0.3 \leq p_T \leq 20$ & \\
\bottomrule
\end{tabular}
\caption{The~used datasets and their kinematic properties (including the center-of-mass energy, the~rapidity window and the transverse momentum range) used for fitting, with the reference for the corresponding article.}
\label{tab:kine}
\end{center}
\end{table}

\vspace{-16pt}

\bibliographystyle{mdpi}

\renewcommand\bibname{References}



\end{document}